\documentclass [12pt]{article}
\usepackage{graphicx,amssymb,amsfonts,latexsym,times,amsmath,amsthm}
\usepackage{epsfig}
\begin{document}

\title{Randomized migration processes between two epidemic centers}

\date{}

\maketitle

\centerline{\bf Igor Sazonov$^a$\footnote{Corresponding author. E-mail: i.sazonov@swansea.ac.uk} and Mark Kelbert$^{b,c}$} 

\centerline{$^a$ College of Engineering, Swansea University, Singleton Park, SA2 8PP, U.K.}

\centerline{$^b$ National Research University Higher School of Economics, Moscow, R.F.} 

\centerline{$^c$ Department of Mathematics, Swansea University, Singleton Park, SA2 8PP, U.K.} 

\begin{abstract}
Epidemic dynamics in a stochastic network of interacting epidemic centers is considered.
The epidemic and migration processes are modelled by Markov's chains. Explicit formulas for probability distribution of the migration process are derived.
Dependence of outbreak parameters on initial parameters, population, coupling parameters is examined analytically and numerically. The mean field approximation for a general migration process is derived. An approximate method allowing computation of statistical moments for networks with highly populated centres is proposed and tested numerically.
\end{abstract}



\section*{Introduction}

Epidemic outbreak in a populated center or in a network of populated centra
develops stochastically due to random interaction between individuals inside
a center and due to a random migration between centers of the network.
Conventionally, an epidemic outbreak in a highly populated center is
described by deterministic processes in according with Law of Large Numbers
(LLN) \cite{Ross}. The mean field approximation (hydrodynamic limits) of the
appropriate statistical models establishes the basic relation of stochastic
description to the dynamical equations, say the SIR model
(susceptible/infected/ removed) and its more sophisticated modifications
(SEIR, SIS, MSIR, etc.)

However, there are two important cases when stochastic effects are
essential. Firstly, it is obviously important when the populations in
centers are not large. The second less obvious scenario can occur at the
initial stage of outbreak when the number of infectives is small, then the
discreteness of population can essentially affect the dynamics of the
outbreak making it stochastic. For an isolated center, this case is
thoroughly studied in \cite{SKG11c}. Analysis of a network of interacting
epidemic centers requires an account of migration fluxes between them.

So, if the initial number of infectives triggering the outbreak in a particular
populated center is small (that is typical for many outbreaks) than the LLN
fails at least at initial stage until the number of infectives is large
enough. For this reason the observed number of infectives can be
significantly different from the prediction of a deterministic model, i.e.
the standard deviation of the number of infectives and the peak outbreak
time can be wide even in highly populated center or a network of such
centers.

In principle, the probability density function (PDF), its standard
deviation, and other important characteristics for the outbreak forecast
could be determined by a direct numerical simulation. However, this
simulation may be very costly and require too much CPU time even for
modern computers. Our goal is to develop a technique for a rough estimation
of the outbreak statistical characteristics skipping such huge computation
by applying some perturbation methods.

Our toolkit is the so-called small initial contagion (SIC) approximation for
the case of a large populated center when initial number of infectives is
small, cf. \cite{SKG11c} in the case of a single populated center. For a
network of highly populated SIR centers in the framework of deterministic
model such technique is described in \cite{SKG08,SKG11a}. In this paper we
develop a stochastic version of SIC approach based on the assumption that in
the real epidemic centra the number of infectives triggering an outbreak is
still small. In a sense, the SIC approach looks like a key to solving
cumbersome epidemic networks.

In \cite{SKG11c}, a randomized analogue of the standard SIR model is
considered. In the SIC approximation, one distinguishes two linked stages of
epidemic evolution. At the stage 1 of initial contamination the number of
infectives is small and a discrete formulation is vital. At this stage the
system is randomized, and governed by stochastic equations. At the stage 2
of developed outbreak when the numbers of individuals in all the components
are large the LLN works and the standard deterministic SIR model can
describe the outbreak process accurately enough. Therefore it is natural to
consider a deterministic system with randomized initial conditions linked to
the stochastic stage~1. 
The statistical characteristics of
the complete model are obtained by applying deterministic equations with
random initial conditions using the matching asymptotic expansions technique
(cf. \cite{Nayfeh}). In contrast to the traditional technique, the
asymptotic approximation of a randomized evolution (at a brief initial
period) is matched with a deterministic evolution with randomized initial
conditions (for all other times). Nevertheless, as in the traditional
approach, we match the approximations at some intermediate time $t$ in the
interval where the both approximations are valid and which drops out from
the final results (cf. \cite{Nayfeh}).

The Markov chain (MC) describing the randomized SIR model has been studied
previously, e.g., in \cite{DG} where a partial differential equation (PDE)
for the moment generating function was derived. We develop a similar
approach for a pair of linked centers and obtain approximate formulae for
their main statistical characteristics. The results of large-scale numerical
simulation are in a good agreement with the appropriate models.

The paper is organized as follows. In Section~\ref{sec:MC} we introduce a MC model describing randomized
epidemic outbreak in two populated centers coupled by a random migration of
all types of species. Here we consider convergence of this model to a
deterministic mean-field model proposed in \cite{arXiv}.
In Section~\ref{sec:migration} we describe a model of random migration
between two interacting SIR centers taking place before the outbreak
started to determine all the initial conditions for the MC
model. Migration between centers is also described as a Markov chain. Here
we derive the Master/Kolmogorov's equations for the probability generating
functions (PGF) and solve them analytically. This analysis confirms the
diffusion-like model of migration heuristically proposed in \cite{arXiv}.
In Section~\ref{sec:numerics} the numerical algorithm for direct solving the
MC model is described, the dependence of outbreak parameters on the population size, initial number of
infectives and migration parameters are presented and discussed.
In Section~\ref{sec:SIC} the generalization of the MC model on an arbitrary
network of the epidemic centers is studied, and difficulties of direct
numerical modelling are estimated. Here the simplified model which looks
relevant for a network of a highly populated centers is proposed.
In Discussions we make a comparison with some previously considered models
and outline the perspective of the future development.

\section{Randomized model of two SIR centra interaction}

\label{sec:MC}

Consider two populated nodes, 1 and 2, with populations $N_{1}$ and $N_{2}$,
respectively. Let $S_{n}(t)$, $I_{n}(t)$, $R_{n}(t)$ be the numbers of host
susceptibles, infectives and removed, respectively, in node~$n$ at time $t$.
Let $S_{mn}(t)$, $I_{mn}(t)$, $R_{mn}(t)$ be numbers of guest susceptibles,
infectives and removed, respectively, in node~$n$ migrated from node $m$ at
time $t$. Note that in the standard SIR model, removed individuals do not
interact with other species, do not affect dynamics of susceptibles and
infectives, and can be omitted from consideration \cite{DG,Mollison95,Murray}.

Assume that the populations in every node is completely mixed, the
contamination rate $\beta _{n}$ of a susceptible individual in node $n$ at
time interval $[t,t+\mathrm{d}t]$ is proportional to the number of all
infectives in node $n$: host infectives $I_{n}$ at time $t$ plus guest
infectives $I_{mn}$ immigrated from node $m$. Next, every infective in node $%
n$ can be removed with probability rate $\alpha _{n}$ (cf.\ \cite{SKG11c}).

The model is described by the migration rate $\gamma _{nm}$ from node $n$ to
node $m$ and return rate $\delta _{mn}$ for a guest individual to return to
his host node, they may be different for different species, i.e.\ we
specify the migration process for susceptibles by parameters $\gamma _{nm}^{S}$,%
$\delta _{mn}^{S}$ and for infectives---$\gamma _{nm}^{I}$,$\delta _{mn}^{I}$
(cf.\ \cite{arXiv}). Obviously, return rate of the host node should be higher than the migration
rate to a neighbour node, i.e., $\gamma _{mn}^{I}<\delta _{nm}^{I},\gamma
_{mn}^{S}<\delta _{nm}^{S}$.

Taking into account all above, we model a network of two SIR centers interacting due to migration of individuals between them a Markov chain (MC) which full description is given in Table~1.

If $I_{0}$ infectives appear in center~1 at time $t=0$ than the initial
conditions for this process are%
\begin{equation}
\begin{array}{@{}l@{\;}l@{\;}l@{\;}l}
I_{1}=I_{0}, & S_{1}=N_{1}-I_{0}-S_{12}^{0}, & I_{12}=0, & S_{12}=S_{12}^{0},
\\
I_{2}=0, & S_{2}=N_{2}-S_{21}^{0}, & I_{21}=0, & S_{21}=S_{21}^{0}.%
\end{array}
\label{eq:ini}
\end{equation}

Here the initial numbers of guest susceptibles $S_{12}^{0}$ and $S_{21}^{0}$
are random and determined by migration processes between centers taking
place before appearing of a single infective. In Section~\ref{sec:migration}
we determine this distribution (which turns out to be binomial) by
considering pure migration processes taking place before the outbreak: see
Eq.~(\ref{eq:S0-distribution}) below. Mean values for $S_{12}^{0}$ and $%
S_{21}^{0}$ are given by (\ref{eq:S12:mean}).

Numerical simulation based on this model is presented and discussed in
Section~\ref{sec:numerics}. Analytical approach via Master/Kolmogorov
equations (see \cite{SKG11c}) is too cumbersome, their analysis and solution
in the general case is rather complicated, so we will develop reasonable
approximations.
\begin{table*}[!h]
\caption{Markov's chain for two coupled SIR nodes}
\label{tab:MC}\centering
\begin{tabular}{lclcl}
\hline\hline
$i$ & Process \#$i$ & Rate, $\ \nu _{i}$ & Restriction & Description \\
\hline\hline
1 & $%
\begin{array}{l@{\,}l@{\,}l@{\,}l@{\,}l}
S_{1} & \rightarrow & S_{1} & - & 1 \\
I_{1} & \rightarrow & I_{1} & + & 1%
\end{array}%
$ & $\beta _{1}(I_{1}{+}I_{21})S_{1}$ & $I_{1} \leq N_{1}$ & Contamination
in 1 (host) \\ \hline
2 & $%
\begin{array}{l@{\,}l@{\,}l@{\,}l@{\,}l}
S_{21} & \rightarrow & S_{21} & - & 1 \\
I_{21} & \rightarrow & I_{21} & + & 1%
\end{array}%
$ & $\beta _{1}(I_{1}{+}I_{21})S_{21}$ & $I_{21}<N_{2}$ & Contamination in 1
(guest) \\ \hline
3 & $%
\begin{array}{l@{\,}l@{\,}l@{\,}l@{\,}l}
I_{1} & \rightarrow & I_{1} & - & 1%
\end{array}%
$ & $\alpha _{1}I_{1}$ &  & Recovering in 1 (host) \rule[-0.5pc]{0pc}{1.5pc}
\\ \hline
4 & $%
\begin{array}{l@{\,}l@{\,}l@{\,}l@{\,}l}
I_{21} & \rightarrow & I_{21} & - & 1%
\end{array}%
$ & $\alpha _{1}I_{21}$ &  & Recovering in 1 (guest)
\rule[-0.5pc]{0pc}{1.5pc} \\ \hline
5 & $%
\begin{array}{l@{\,}l@{\,}l@{\,}l@{\,}l}
S_{1} & \rightarrow & S_{1} & - & 1 \\
S_{12} & \rightarrow & S_{12} & + & 1%
\end{array}%
$ & $\gamma _{12}^{S}S_{1}$ & $S_{12} \leq N_{1}$ & Migration from 1 to 2 \\
\hline
6 & $%
\begin{array}{l@{\,}l@{\,}l@{\,}l@{\,}l}
I_{1} & \rightarrow & I_{1} & - & 1 \\
I_{12} & \rightarrow & I_{12} & + & 1%
\end{array}%
$ & $\gamma _{12}^{I}I_{1}$ & $I_{12} \leq N_{1}$ & Migration from 1 to 2
\\ \hline
7 & $%
\begin{array}{l@{\,}l@{\,}l@{\,}l@{\,}l}
S_{1} & \rightarrow & S_{1} & + & 1 \\
S_{12} & \rightarrow & S_{12} & - & 1%
\end{array}%
$ & $\delta _{21}^{S}S_{12}$ & $S_{1} \leq N_{1}$ & Return from 2 to 1 \\
\hline
8 & $%
\begin{array}{l@{\,}l@{\,}l@{\,}l@{\,}l}
I_{1} & \rightarrow & I_{1} & + & 1 \\
I_{12} & \rightarrow & I_{12} & - & 1%
\end{array}%
$ & $\delta _{21}^{I}I_{12}$ & $I_{1} \leq N_{1}$ & Return from 2 to 1 \\
\hline\hline
9 & $%
\begin{array}{l@{\,}l@{\,}l@{\,}l@{\,}l}
S_{2} & \rightarrow & S_{2} & - & 1 \\
I_{2} & \rightarrow & I_{2} & + & 1%
\end{array}%
$ & $\beta _{2}(I_{2}{+}I_{12})S_{2}$ & $I_{2} \leq N_{2}$ & Contamination
in 2 (host) \\ \hline
10 & $%
\begin{array}{l@{\,}l@{\,}l@{\,}l@{\,}l}
S_{12} & \rightarrow & S_{12} & - & 1 \\
I_{12} & \rightarrow & I_{12} & + & 1%
\end{array}%
$ & $\beta _{2}(I_{2}{+}I_{12})S_{12}$ & $I_{12} \leq N_{1}$ & Contamination
in 2 (guest) \\ \hline
11 & $%
\begin{array}{l@{\,}l@{\,}l@{\,}l@{\,}l}
I_{2} & \rightarrow & I_{2} & - & 1%
\end{array}%
$ & $\alpha _{2}I_{2}$ &  & Recovering in 2 (host) \rule[-0.5pc]{0pc}{1.5pc}
\\ \hline
12 & $%
\begin{array}{l@{\,}l@{\,}l@{\,}l@{\,}l}
I_{12} & \rightarrow & I_{12} & - & 1%
\end{array}%
$ & $\alpha _{2}I_{12}$ &  & Recovering in 2 (guest)
\rule[-0.5pc]{0pc}{1.5pc} \\ \hline
13 & $%
\begin{array}{l@{\,}l@{\,}l@{\,}l@{\,}l}
S_{2} & \rightarrow & S_{2} & - & 1 \\
S_{21} & \rightarrow & S_{21} & + & 1%
\end{array}%
$ & $\gamma _{21}^{S}S_{2}$ & $S_{21} \leq N_{2}$ & Migration from 2 to 1 \\
\hline
14 & $%
\begin{array}{l@{\,}l@{\,}l@{\,}l@{\,}l}
I_{2} & \rightarrow & I_{2} & - & 1 \\
I_{21} & \rightarrow & I_{21} & + & 1%
\end{array}%
$ & $\gamma _{21}^{I}I_{2}$ & $I_{21} \leq N_{2}$ & Migration from 2 to 1 \\
\hline
15 & $%
\begin{array}{l@{\,}l@{\,}l@{\,}l@{\,}l}
S_{2} & \rightarrow & S_{2} & + & 1 \\
S_{21} & \rightarrow & S_{21} & - & 1%
\end{array}%
$ & $\delta _{12}^{S}S_{21}$ & $S_{2} \leq N_{2}$ & Return from 1 to 2 \\
\hline
16 & $%
\begin{array}{l@{\,}l@{\,}l@{\,}l@{\,}l}
I_{2} & \rightarrow & I_{2} & + & 1 \\
I_{21} & \rightarrow & I_{21} & - & 1%
\end{array}%
$ & $\delta _{12}^{I}I_{21}$ & $I_{2} \leq N_{2}$ & Return from 1 to 2 \\
\hline\hline
\end{tabular}%
\end{table*}



\noindent \textbf{Proposition 1.} The scaled Markov chain (MC) $\left\{
I_{n}^{\ast }(t),S_{n}^{\ast }(t),I_{nm}^{\ast }(t),S_{nm}^{\ast
}(t)\right\} $ ($n=1,2$, $m=2,1$)%
\begin{equation}
\begin{array}{r@{\;}c@{\;}lr@{\;}c@{\;}l}
I_{n}^{\ast }(t) & = & \Lambda ^{-1}I_{n}(t), & S_{n}^{\ast }(t) & = &
\Lambda ^{-1}S_{n}(t), \\
I_{nm}^{\ast }(t) & = & \Lambda ^{-1}I_{nm}(t), & S_{nm}^{\ast }(t) & = &
\Lambda ^{-1}S_{nm}(t)%
\end{array}
\label{eq:scale}
\end{equation}%
in a populations of sizes $\Lambda N_{n},\Lambda N_{m}$ related with process
$\{ I_{n}(t),S_{n}(t),I_{nm}(t),$ $S_{nm}(t)\} $ defined in Table 1
and subjected to initial conditions (\ref{eq:ini}) by scaling the transition
rates $\beta _{n}\rightarrow \Lambda ^{-1}\beta _{n}$, and scaling of
initial conditions as%
\begin{equation}
\begin{array}{r@{\;}c@{\;}lr@{\;}c@{\;}l}
I_{1}(0) & = & \Lambda I_{0}, & S_{1}(0) & = & \Lambda
(N_{1}-I_{0}-S_{12}^{0}), \\
I_{2}(0) & = & 0, & S_{2}(0) & = & \Lambda \left( N_{2}-S_{21}^{0}\right) ,
\\
I_{12}(0) & = & 0, & S_{12} & = & \Lambda S_{12}^{0}, \\
I_{21}(0) & = & 0, & S_{21} & = & \Lambda S_{21}^{0}%
\end{array}
\label{eq:ini:scale}
\end{equation}
where the PDFs of independent random variables $S_{12}^{0}$ and $S_{21}^{0}$ have binomial distributions (\ref{eq:S0-distribution}), converges in distribution as $%
\Lambda \rightarrow \infty $ to the deterministic functions $\left\{ \hat{I}%
_{n}(t),\hat{S}_{n}(t),\hat{I}_{nm}(t),\hat{S}_{nm}(t)\right\} $ satisfying
ODEs introduced in \cite{arXiv}
\begin{eqnarray}
\textstyle{\frac{\mathrm{d}}{\mathrm{d}t}}\hat{S}_{n} &=&-\beta _{n}\hat{S}%
_{n}(\hat{I}_{n}+\hat{I}_{mn})-\gamma _{nm}^{S}\hat{S}_{n}+\delta _{mn}^{S}\hat{S}_{nm}
\label{ODE:Sn} \\
\textstyle{\frac{\mathrm{d}}{\mathrm{d}t}}\hat{I}_{n} &=&\beta _{n}\hat{S}%
_{n}(\hat{I}_{n}+\hat{I}_{mn})-\alpha _{n}\hat{I}_{n}  -\gamma _{nm}^{I}\hat{I}_{n}+\delta _{mn}^{I}\hat{I}_{nm}
\label{ODE:In} \\
\textstyle{\frac{\mathrm{d}}{\mathrm{d}t}}\hat{S}_{mn} &=&-\beta _{n}\hat{S}%
_{mn}(\hat{I}_{n}+\hat{I}_{mn})  +\gamma _{mn}^{S}\hat{S}_{m}-\delta _{nm}^{S}\hat{S}_{mn}
\label{ODE:Smn} \\
\textstyle{\frac{\mathrm{d}}{\mathrm{d}t}}\hat{I}_{mn} &=&\beta _{n}\hat{S}%
_{mn}(\hat{I}_{n}+\hat{I}_{mn})-\alpha \hat{I}_{mn}  +\gamma _{mn}^{I}\hat{I}_{m}-\delta _{nm}^{I}\hat{I}_{mn}
\label{ODE:Imn}
\end{eqnarray}%
subjected to the initial conditions%
\begin{equation}
\begin{array}{r@{\;}c@{\;}lr@{\;}c@{\;}l}
\hat{I}_{1}(0) & = & I_{0}, & \hat{S}_{1}(0) & = & N_{1}-I_{0}-\bar{S}%
_{12}^{0}, \\
\hat{I}_{2}(0) & = & 0, & \hat{S}_{2}(0) & = & N_{2}-\bar{S}_{21}^{0}, \\
\hat{I}_{12}(0) & = & 0, & \hat{S}_{12} & = & \bar{S}_{12}^{0}, \\
\hat{I}_{21}(0) & = & 0, & \hat{S}_{21} & = & \bar{S}_{21}^{0}%
\end{array}
\label{eq:ODE:ini}
\end{equation}%
where%
\begin{equation}
\bar{S}_{12}^{0}=\frac{\gamma _{12}^{S}N_{1}}{\gamma _{12}^{S}+\delta
_{21}^{S}},\ \bar{S}_{21}^{0}=\frac{\gamma _{21}^{S}N_{2}}{\gamma
_{21}^{S}+\delta _{12}^{S}}  \label{eq:S12:mean}
\end{equation}%
\bigskip

In fact, equations (\ref{ODE:Sn})--(\ref{ODE:Imn}) can be derived
phenomenologically: if the number of species is large enough, its
change by one or by few can be considered as infinitesimally small. For
example, the number of infectives $I_{1}$ can increase due to process \#1
and \#8 with rates $\beta _{1}(I_{1}{+}I_{21})S_{1}$ and $\delta
_{21}^{I}I_{12}$, respectively, or decrease due to process \#3 and \#6 with
rates $\alpha _{1}I_{1}$ and $\gamma _{12}^{I}I_{1}$, respectively. Therefore
the rate of $\mathrm{d}I_{1}$ in time
interval $\mathrm{d}t$ can be estimated as $\mathrm{d}I_{1}=\left[ \beta
_{1}(I_{1}{+}I_{21})S_{1}+\delta _{21}^{I}I_{12}\right] \!\mathrm{d}t-\left[
\alpha _{1}I_{1}+\gamma _{12}^{I}I_{1}\right] \!\mathrm{d}t$, that gives
Eq.~(\ref{ODE:In}) for $n=1,m=2$. The same holds for all other species. This
argument can be made rigorous with the help of Law of Large Numbers (LLN).
Then it indicates that the mean values of the variables converge to the
mean-field (hydrodynamic) limit.

The more delicate question is to establish the convergence in probability.
Sketch of the proof is given in Appendix.

\section{Random migration of non-contaminating species}

\label{sec:migration}

In order to elaborate distributions for $S_{nm}^{0}$ taking place at the
very beginning of the outbreak we study the pure migration process setting $%
I_{1}\equiv 0,I_{2}\equiv 0$. In this case the MC described in Table~1 can
be split into two independent processes: $S_{1}\leftrightarrow
S_{12}=N_{1}-S_{1}$, and\ $S_{2}\leftrightarrow S_{21}=N_{2}-S_{2}$. For
each of them we have the following MC in terms of a single random variable $%
S_{n}$, $n=1,2$:%
\begin{equation}
\begin{tabular}{ll}
\hline\hline
Process & Rate \\ \hline\hline
$S_{n}\rightarrow S_{n}-1$ & $\gamma _{nm}^{S}S_{n}$ \\
$S_{n}\rightarrow S_{n}+1$ & $\delta _{mn}^{S}(N_{n}-S_{n})$ \\ \hline\hline
\end{tabular}
\label{MC:migration}
\end{equation}%
\bigskip

Let $P_{k}(t)=\mathbb{P}(S_{nm}(t){=}k)\equiv \mathbb{P}(S_{n}(t){=}N_{n}-k)$
be the probability distribution in node $m$ at instant $t$. Then
Masters/Kolmogorov's equations take the form
\begin{equation}
\frac{\mathrm{d}}{\mathrm{d}t}P_{k}  =  \gamma \left( \left\lceil
N-k+1\right\rceil _{0}^{N}\right) P_{k-1}-\gamma (N-k)P_{k} +\delta \left( \left\lceil k+1\right\rceil
_{0}^{N}\right) P_{k+1}-\delta kP_{k}%
\label{eq:master-migration}
\end{equation}%
where $0\leq k\leq N$; for the sake of simplicity we temporary set $\gamma
=\gamma _{nm}^{S}$, $\delta =\delta _{mn}^{S}$, $N=N_{n}$. Here the
following notation is introduced to write equations for $k=0$ and $k=N$ in
the same form as others
\begin{equation}
\left\lceil k\right\rceil _{0}^{N}=\left\{
\begin{array}{cc}
k, & 0\leq k\leq N \\
0, & \text{otherwise.}%
\end{array}%
\right.  \label{eq:notation}
\end{equation}%
Equations (\ref{eq:master-migration}) implies the following PDE
\begin{equation}
G_{t}=\left( 1-z\right) \left[ \left( -\gamma z-\delta \right) G_{z}+\gamma
N\,G\,\right]   \label{eq:PDE-migr}
\end{equation}%
for the probability generating function (PGF)
\begin{equation}
G(z,t)=\sum\limits_{k=0}^{N}z^{k}P_{k}(t).  \label{eq:pgf}
\end{equation}%
The initial condition $P_{0}(0)=1,P_{k>1}(0)=0$ implies
\begin{equation}
G(z,0)=1.  \label{eq:G:ini}
\end{equation}

The solution to problem (\ref{eq:PDE-migr})--(\ref{eq:G:ini}) can be found
explicitly
\begin{equation}
G(z,t)=\left[ \frac{\left( \gamma z+\delta \right) -\gamma (z-1)e^{-\left(
\gamma +\delta \right) t}}{\gamma +\delta }\right] ^{N}.  \label{eq: PDE-sol}
\end{equation}%
Now one can easily calculate all the moments of distribution $\{P_{k}(t)\}$,
say
\begin{eqnarray}
\mathbb{E}S(t) & \equiv & \mu _{1}(t)=G_{z}(1,t)=N\varepsilon \left[
1-e^{-t/\tau }\right]%
\label{eq:mu1:migration} \\
\mathrm{var}(S(t)) & \equiv & \mu _{2}(t)=G_{zz}(1,t)+\mu _{1}-\mu _{1}^{2}
=\mu _{1}(t)\left[ \varepsilon e^{-t/\tau
}+(1-\varepsilon )\right]%
\label{eq:mu2:migration}
\end{eqnarray}%
where $\varepsilon =\gamma /(\gamma +\delta )$, $\tau =1/(\gamma +\delta )$.%

If the migration process lasts long enough before the outbreak starts then
the PGF takes its limiting form for $t\rightarrow \infty $%
\begin{equation*}
G(z,\infty )=\left( \varepsilon z+(1-\varepsilon )\right) ^{N}
\end{equation*}%
which is the MGF for a binomial distribution:%
\begin{equation}
\mathbb{P}(S_{nm}^{0}=k)=\binom{N_{n}}{k}\ \left( \varepsilon
_{nm}^{S}\right) ^{k}(1-\varepsilon _{nm}^{S})^{N_{n}-k}.
\label{eq:S0-distribution}
\end{equation}%
From here on we return to the indexed notation%
\begin{equation}
\varepsilon _{nm}^{S,I}=\frac{\gamma _{nm}^{S,I}}{\gamma _{nm}^{S,I}+\delta
_{mn}^{S,I}}.  \label{epsSI}
\end{equation}

This distribution has the following first two moments%
\begin{eqnarray}
\bar{S}_{nm}^{0} &\equiv &\mathbb{E}S_{nm}^{0}=\mu _{1}(\infty )=\varepsilon
_{nm}^{S}N_{n}  \label{eq:mu1(infty)} \\
\mathrm{var}(S_{nm}^{0}) &=&\mu _{2}(\infty )=N_{n}\varepsilon
_{nm}^{S}(1-\varepsilon _{nm}^{S}).  \label{eq:mu2(infty)}
\end{eqnarray}%
The relative standard deviation (i.e. for the process $X=S_{nm}/\bar{S}%
_{nm}^{0})$ decays as $N_{n}^{-1/2}$:
\begin{equation}
\sigma _{S_{nm}^{0}/\bar{S}_{nm}^{0}}=\frac{\sqrt{\mu _{2}(\infty )}}{\mu
_{1}(\infty )}=\sqrt{\frac{(1-\varepsilon _{nm}^{S})}{\varepsilon
_{nm}^{S}N_{n}}}.  \label{eq:STD:migration}
\end{equation}%
Hence, when $N_{n}\rightarrow \infty $, the migration process tends in
probability to the deterministic limit described in \cite{arXiv}.

Thus, in the MC model defined by Table~\ref{tab:MC} the initial conditions $%
S_{12}^{0}$ and $S_{21}^{0}$ can be selected randomly with the binomial
distribution (\ref{eq:S0-distribution}) or approximated by a Gaussian
function if $N_{n}$ is large enough.

Parameter $\varepsilon _{nm}^{S}:=\gamma _{nm}^{S}/(\gamma _{nm}^{S}+\delta
_{nm}^{S})=\bar{S}_{nm}^{0}/N_{n}$ indicates the mean share of individuals
from node $n$ migrated to node $m$. Obviously this share in average should
be small for highly populated centers: say, half population of a city hardly
can be on a visit into another city for an essential time. Parameter $%
\varepsilon _{nm}^{S}$ can be treated as a coupling parameter characterizing
how intensive are migration fluxes between populated centers. Analogous
fluxes of infectives hardly are more intensive, therefore $\varepsilon
_{nm}^{I}=\gamma _{nm}^{I}/(\gamma _{nm}^{I}+\delta _{nm}^{I})\leq
\varepsilon _{nm}^{S}$. So, for highly populated centers, the following
inequality should be kept
\begin{equation}
\varepsilon _{nm}^{S,I}\ll 1\Longleftrightarrow \gamma _{mn}^{S,I}\ll \delta
_{nm}^{S,I}.  \label{eq:e<<1}
\end{equation}

The second important characteristic of migration process is \ the
characteristic migration time \ $\tau _{nm}^{S,I}=1/(\gamma
_{nm}^{S,I}+\delta _{nm}^{S,I})$. From (\ref{eq:mu1:migration}) one can see
that it indicates how soon the dynamic equilibrium established after the
migration process started or the population changes suddenly. Both pairs of
parameters: $\left\{ \gamma ,\delta \right\} $ and $\{\varepsilon ,\tau \}$
are uniquely related.

\section{Direct numerical simulation of two interacting SIR centra}

\label{sec:numerics}

\subsection{Numerical scheme}

In the numerical simulation of the randomized SIR model the time interval
was divided into small steps $\Delta t$ such that the sum of all rates from
Table~1 
multiplied by $\Delta t$ is essentially less than 1:
\begin{equation}
\max_{t}\left\{ \nu _{\Sigma }(t)\right\} \Delta t\ll 1\Longrightarrow
\Delta t=\min \left\{ P_{t}/\nu _{\Sigma }(t)\right\}  \label{eq:rate:sum}
\end{equation}%
where $P_{t}$ is the admitted threshold, say, $P_{t}=0.1$\thinspace .

Probability that at least one event occurs in unit of time is bounded by the
sum of rates of all the processes $%
\begin{array}{ccc}
\nu _{\Sigma }(t) & = & \sum_{i=1}^{16}\nu _{i}%
\end{array}%
$:
\begin{eqnarray*}
\nu _{\Sigma }(t) & = & \beta _{1}\left( I_{1}+I_{21}\right) \left(
S_{1}+S_{21}\right)  + \beta _{2}\left( I_{2}+I_{12}\right) \left( S_{2}+S_{12}\right) \\
& + & \alpha _{2}(I_{2}+I_{12})+\alpha _{1}(I_{1}+I_{21}) \\
& + & \gamma _{12}^{S}S_{1}+\gamma _{12}^{I}I_{1}+\delta
_{21}^{S}S_{12}+\delta _{21}^{I}I_{12}
+ \gamma _{21}^{S}S_{2}+\gamma _{21}^{I}I_{2}+\delta
_{21}^{S}S_{12}+\delta _{12}^{I}I_{21}%
\end{eqnarray*}%
In this relation we majorize $I_{1},S_{1}\leq N_{1}$, $I_{2},S_{2}\leq N_{2}$%
, $I_{12},S_{12}\leq N_{1}$, $I_{21},S_{21}\leq N_{2}$, then
\begin{eqnarray*}
\max (\nu _{\Sigma }) & = & (\beta _{1}+\beta _{2})(N_{1}+N_{2})^{2}
 + (\alpha _{1}+\alpha _{2})(N_{1}+N_{2}) \\
& + & (\gamma _{12}^{S}+\gamma _{12}^{I}+\delta _{21}^{I}+\delta
_{21}^{S})N_{1}
 + (\gamma _{21}^{S}+\gamma _{21}^{I}+\delta _{12}^{I}+\delta
_{12}^{S})N_{2}  \\
& + &
\gamma _{12}^{S}S_{1}+\gamma _{12}^{I}I_{1}+\delta
_{21}^{S}S_{12}+\delta _{21}^{I}I_{12}
\end{eqnarray*}%
Actually this value overestimates the really occurred total rate
significantly as it is very improbable that numbers of guest susceptibles and infectives  in a highly populated center essentially exceeds values $%
\varepsilon _{nm}^{S}N_{n}$ and $\varepsilon _{nm}^{I}N_{n}$, respectively,
where $\varepsilon _{nm}^{S,I}$ is defined in (\ref{eq:S0-distribution}), in
virtue of (\ref{eq:e<<1}). For this reason we can account that $%
S_{nm}\lesssim \varepsilon _{nm}^{S}N_{n}$ and $I_{nm}\lesssim \varepsilon
_{nm}^{I}N_{n}$ (and also use the rigorous inequalities $I_{n}S_{n}\leq
\frac{1}{4}N_{n}^{2}$). Then we obtain the more realistic estimation:%
\begin{eqnarray*}
\max \left( \nu _{\Sigma }\right) & \simeq & \textstyle{%
\frac{1}{4}}\beta _{1}N_{1}^{2}+\textstyle{\frac{1}{4}}\beta _{2}N_{2}^{2}
\\&+ &\beta _{1}\left( \varepsilon _{21}^{I}+\varepsilon _{21}^{S}\right)
N_{2}N_{1}
+ \beta _{2}\left( \varepsilon _{12}^{I}+\varepsilon _{12}^{S}\right)
N_{2}N_{1} \\
& + & \alpha _{1}\left( N_{1}+\varepsilon _{21}^{I}N_{2}\right) +\alpha
_{2}\left( N_{2}+\varepsilon _{12}^{I}N_{1}\right) \\
& + & \left( \gamma _{12}^{S}+\gamma _{12}^{I}\right) N_{1}+\left(
\varepsilon _{21}^{S}\delta _{12}^{S}+\varepsilon _{21}^{I}\delta
_{12}^{I}\right) N_{1} \\
& + & \left( \gamma _{21}^{S}+\gamma _{21}^{I}\right) N_{2}+\left(
\varepsilon _{12}^{S}\delta _{21}^{S}+\varepsilon _{12}^{I}\delta
_{21}^{I}\right) N_{2} \\
& + & \beta _{1}\varepsilon _{21}^{I}\varepsilon _{21}^{S}N_{2}N_{1}+\beta
_{2}\varepsilon _{12}^{I}\varepsilon _{12}^{S}N_{2}N_{1}%
\end{eqnarray*}

The following numerical scheme is used:

\begin{enumerate}
\item Assign the initial values to 8 variables%
\begin{equation*}
\begin{array}{@{}l@{\;}l@{\;}l@{\;}l}
I_{1}=I_{0}, & S_{1}=N_{1}{-}I_{0}{-}S_{12}^{0}, & I_{12}=0, &
S_{12}=S_{12}^{0}, \\
I_{2}=0, & S_{2}=N_{2}{-}S_{21}^{0}, & I_{21}=0, & S_{21}=S_{21}^{0}.%
\end{array}%
\end{equation*}
\noindent where $S_{12}^{0},S_{21}^{0}$ are random numbers distributed in
accordance with (\ref{eq:S0-distribution}).

\item Calculate the current rates $\left\{ \nu _{i},i=1,\ldots ,16\right\} $
indicated in Table 1.

\item Calculate the current probability of at least one event occurrence in
accordance with eq.~(\ref{eq:rate:sum}):%
\begin{equation*}
p=\Delta t\,\nu _{\Sigma }(t)\equiv \Delta t\sum_{i=1}^{16}\nu _{i}
\end{equation*}

\item Generate uniformly distributed random number$\ x\in \lbrack 0,1]$.

\item If $x>p$ then no events occur. In this case:

\begin{enumerate}
\item advance at one time step: $t\leftarrow t+\Delta t$ without changing
variables $I_{1},\ldots ,S_{21}$, also $\nu _{1},\ldots ,\nu _{16}$ and $p$.

\item if $t>t_{\max }$ terminate the process, otherwise go to step~4.
\end{enumerate}

If $x\leq p$ then at least one event occurs. In this case:

\begin{enumerate}
\item calculate the intervals $\Delta y_{i}=[\eta _{i-1},\eta _{i}]$, $\eta
_{i}=\sum_{j=1}^{i}\nu _{j}$

\item generate the second random number $y$ uniformly distributed in $%
\lbrack 0,\eta _{16}]$

\item find in which interval $y$ falls.

\item perform the process described in Table 1 with the correspondent rate

\item advance at one time step: $t\leftarrow t+\Delta t$

\item if $t>t_{\max }$ terminate the process, otherwise go to step~2.\bigskip
\end{enumerate}
\end{enumerate}

\subsection{Numerical results}

\label{sec:numerics-results}

For numerical computation a basic model with two identical centers has
parameters:\ $N_{1,2}=10^{4}$, $\alpha _{1,2}=1$, $\mathrm{Ro}_{1,2}=4$, $%
\varepsilon _{1,2}^{I,S}=0.01$, $\tau _{1,2}^{I,S}=5$ where $\mathrm{Ro}%
_{1,2}=\left( \beta _{1,2}/\alpha _{1,2}\right) N_{1,2}$ are reproduction
numbers. The initial number of infectives in the first node $%
I_{0}=i_{0}N_{1}$ where $i_{0}=0.01$ is taken for the basic model.

A few realizations of numerical computation are depicted in Figure~\ref%
{fig:realisation}. Here the time variations of the total number of
infectives $I_{n}^{\Sigma }=I_{n}+I_{mn}$ in every node are shown and
compared with the curves based on integration of the deterministic initial
value problem (\ref{ODE:Sn})--(\ref{eq:S12:mean})

\begin{figure}[tbh]
\centering \includegraphics[width=80mm]{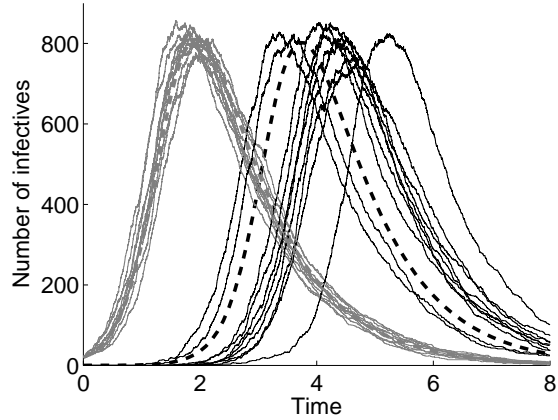} 
\caption{Examples of realizations of two randomized SIR models. The total
number of infectives is plotted in node~1 ($I_{1}^{\Sigma }=I_{1}+I_{21}$)
by thin grey lines and in node~2 ($I_{2}^{\Sigma }=I_{2}+I_{12}$) by thin black lines.
Bold dashed lines indicates the hydrodynamic limit. ($N_1=N_2=2$k, ${\rm Ro}_1={\rm Ro}_2=4$, $\varepsilon=0.01,\tau=5$, $I_0/N_1=0.01$)}
\label{fig:realisation}
\end{figure}

\begin{figure}[!h]
\centering \includegraphics[width=80mm]{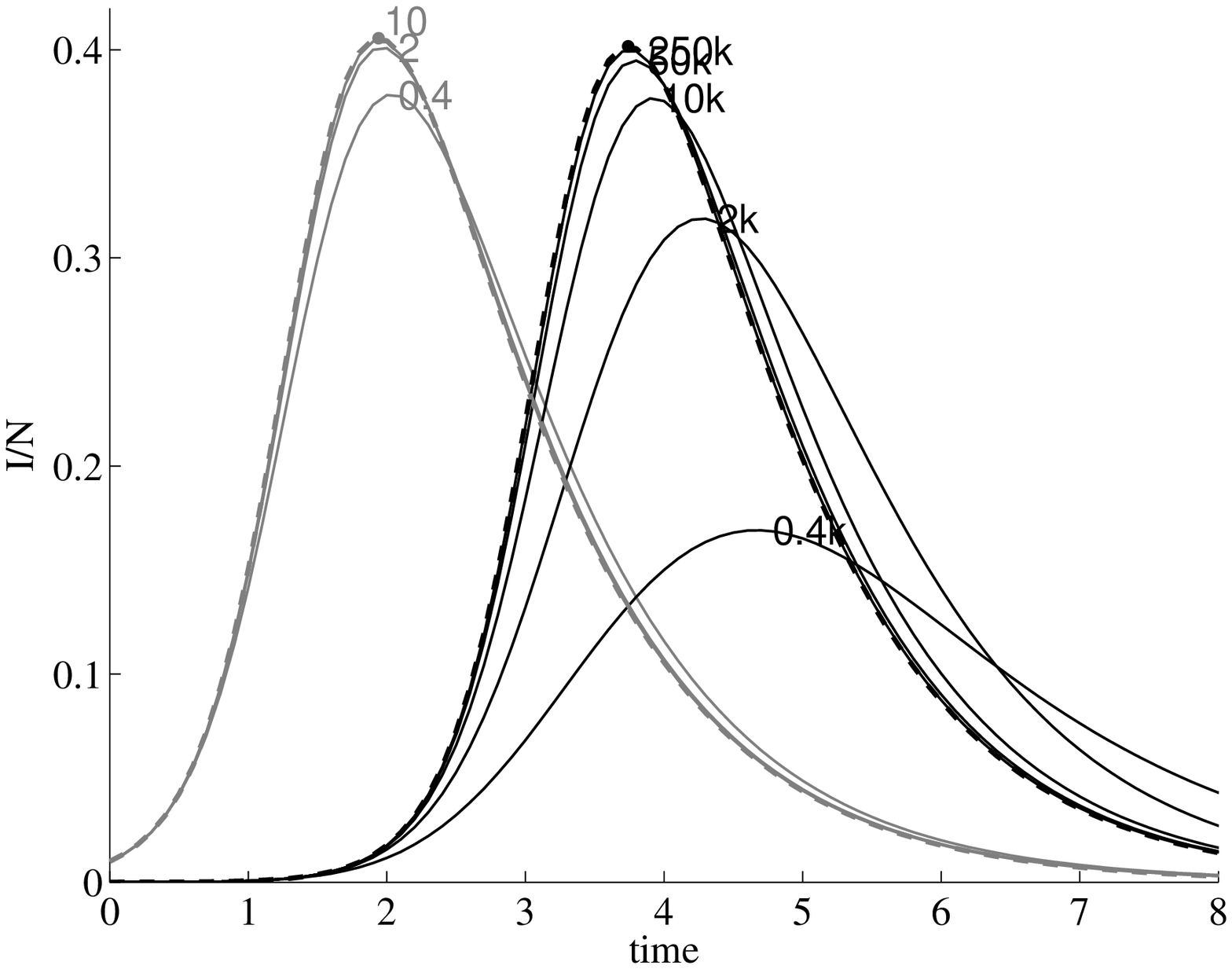} 
\includegraphics[width=80mm]{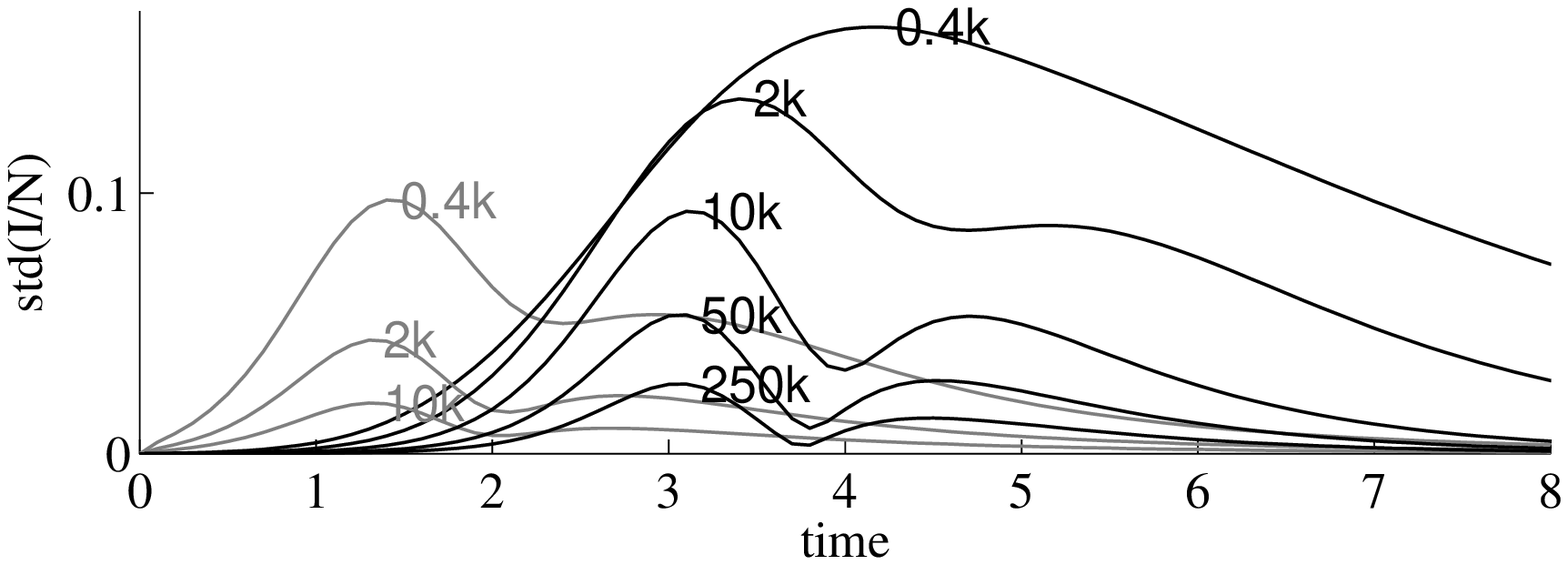}
\caption{Evolution of the mean values for $I^{\Sigma}_m/N_m$ $m=1,2$ (left) 
and their standard deviations (right). 
Grey curves for node~1, black lines
for node~2. Dashed lines indicate the hydrodynamic limit described by eqs.~(%
\protect\ref{ODE:Sn})--(\protect\ref{ODE:Imn}). The node population is
indicated near the top of the correspondent curve.}
\label{fig:I-vs-N}
\end{figure}

In the first set of numerical experiments the total population size
varies from $N_{1}=N_{2}=N=400$ up to 10$^{6}$. Number of realization $L$
was taken $10^{4}$ mainly but the number of realization is taken greater for
small population $N=400$ and $2000$ and smaller for extremely high
populations: 250k and 1000k. The current mean number of total infectives $%
\bar{I}_{1,2}^{\Sigma }(t)$ and standard deviation are computed. The results
of the first set are shown in Figure~\ref{fig:I-vs-N}. Observe that the
mean value of the random process (thin lines) converges to the solution of
correspondent deterministic problem (bold lines). But the convergence is
much slower for node~2: for $N_{1}=10$k the mean trajectory practically
coincides with the deterministic limit, on the other side the same effect in
node 2 requires $N_{2}\sim 250$k.

\begin{figure}[!h]
\centering \includegraphics[width=80mm]{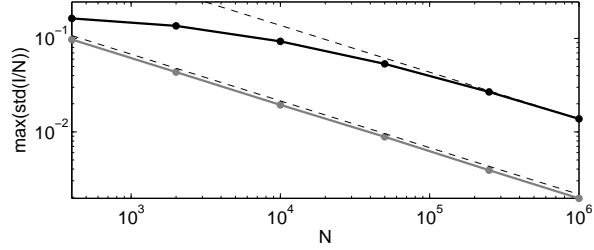} %
\caption{Maximal value of the standard deviation for processes $I_1/N_1$ and
$I_2/N_2$ vs population $N=N_1=N_2$. Black curves for node~1, grey lines for
node~2. The slope of dashed line corresponds to the decay law $N^{-1/2}$.}
\label{fig:stdImax-vs-N}
\end{figure}

The convergence rate is examined in Figure~\ref{fig:stdImax-vs-N}. One
can see that for node~1 the convergence rate almost coincides with $%
O(N^{-1/2})$, as for node~2 the decay rate tends to $O(N^{-1/2})$ only for
sufficiently large population: $N=10^{6}$. Thus for the secondary
contaminated node the account of randomness is essential even if its
population is large provided that the migration parameters $\varepsilon
_{1,2}^{I,S}$ are small (0.01 in this case). Say, if $N_{2}=400$ the
standard deviation exceeds the mean value up to the outbreak time.

\begin{figure}[!h]
\centering \includegraphics[width=80mm]{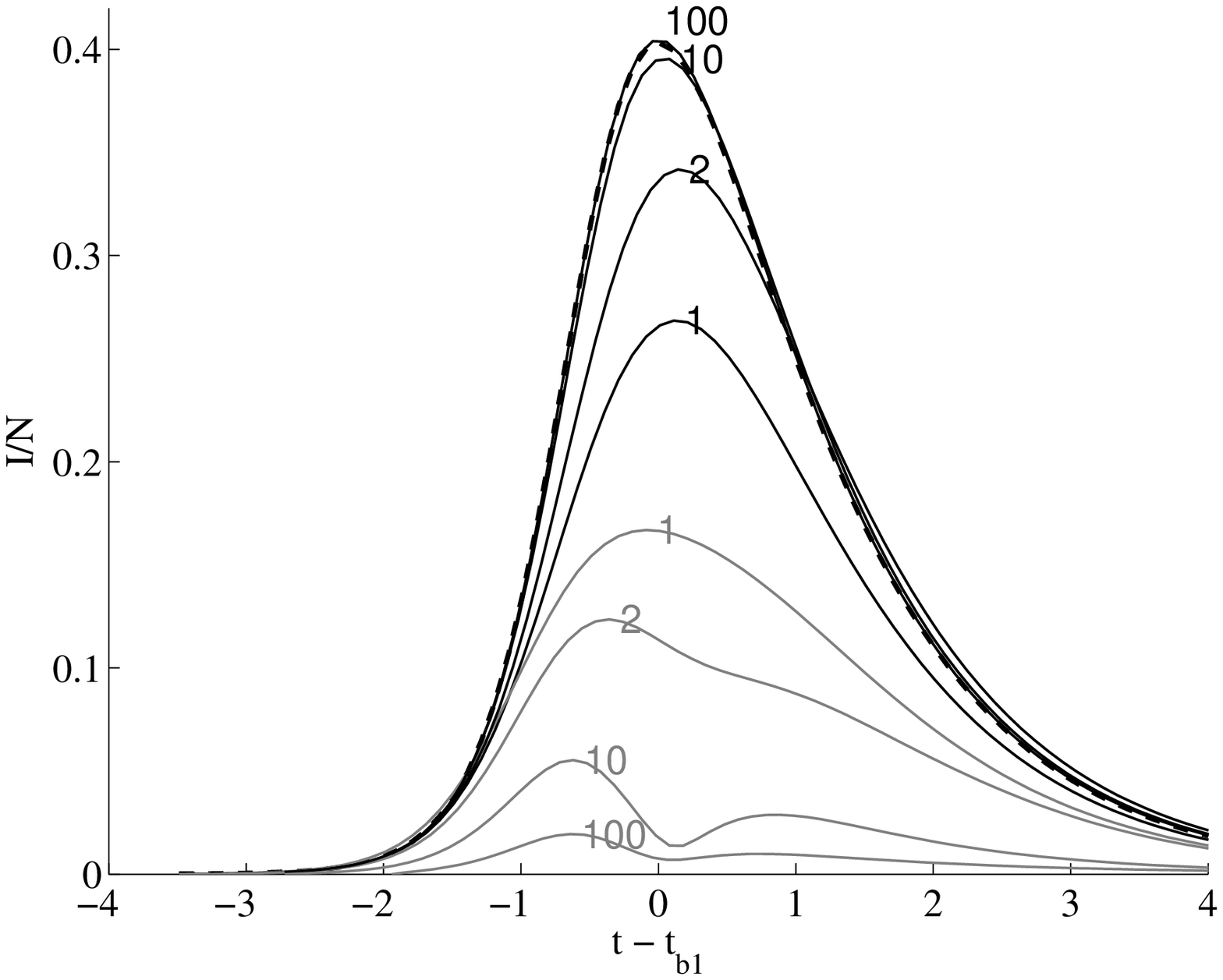} 
\includegraphics[width=80mm]{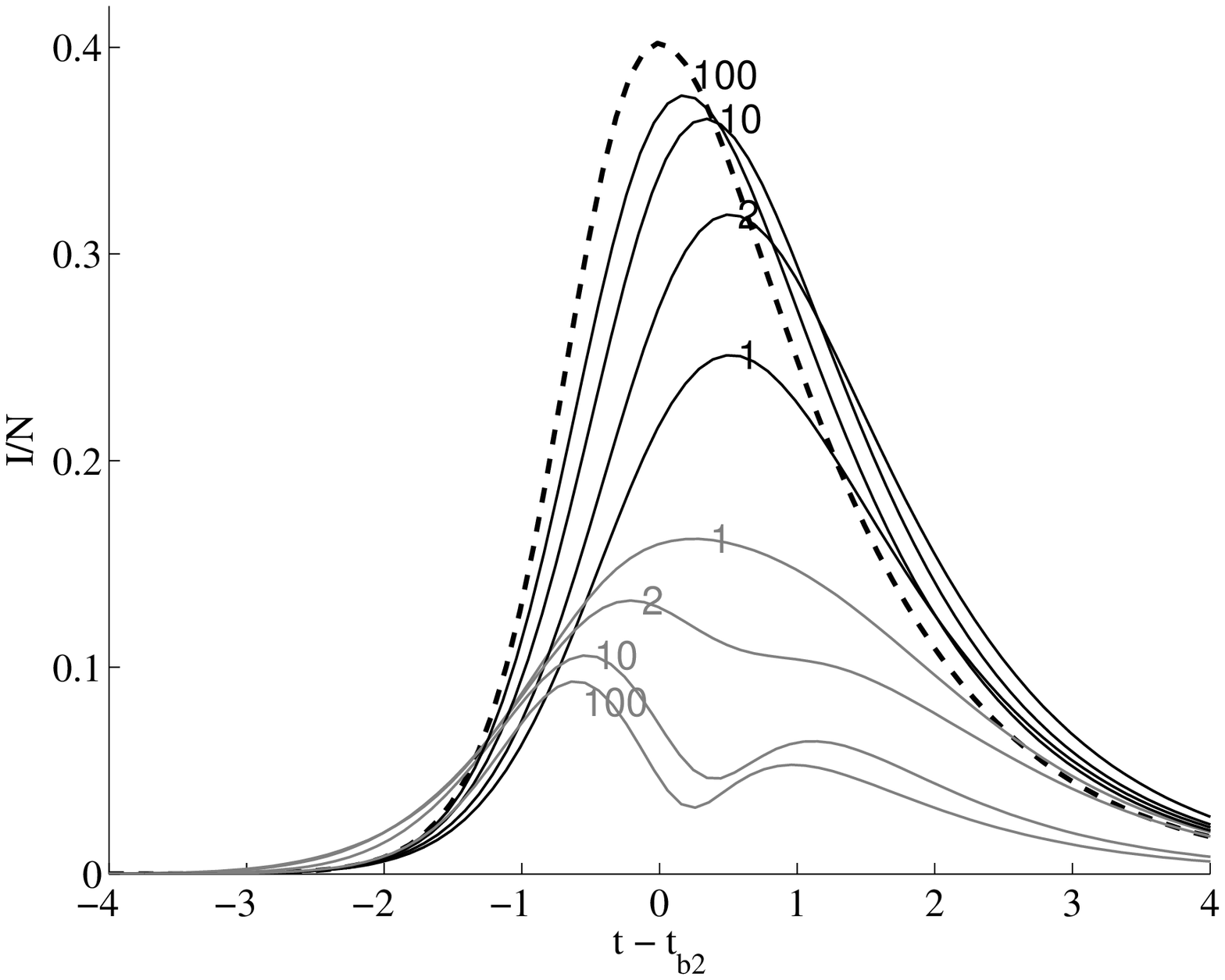}
\caption{Left: Mean values for $I_{1}^{\Sigma }/N_{1}$ (black) and its standard
deviation $\mathrm{std}{I_{1}^{\Sigma }/N_{1}}$ (grey) for different $I_{0} $%
. The initial number of infectives in node~1 is indicated near top of the
corresponding curve. Dashed line indicates the deterministic limiting
solution. Right: the same for $I_{2}^{\Sigma }/N_{2}$.}
\label{fig:I1-vs-I0}
\end{figure}

In the second set of numerical experiments we study dependence of mean
number of infectives and their standard deviation from the initial number of
infectives $I_{0}$ in node 1. It varies from $1$ to $100$ (the share $%
i_{0}=I_{0}/N_{1}$ varies from $10^{-4}$ to $10^{-2}$). The results are
plotted in Figure~\ref{fig:I1-vs-I0}. 
Because the outbreak time depends on initial number of infectives,
the mean-field curves become essentially different. Therefore it is appropriate
to shift the time so that the peak outbreak for different initial condition
is at the same instant, say, $t=0$. Then all the curve are very close to
each other and practically coincide with curve for the limiting solution
introduces in \cite{SKG08, SKG11a}. Observe that the smaller the number of
initial infectives the greater is the standard deviation (std) and the
larger is the difference between the mean curve for randomized process and
the mean-field curve. Also observe that for node~1, the discrepancy of mean
number of infectives from the hydrodynamic limit as well as the standard
deviation monotonically decay with the growth of $I_{0}$. In node~2 the
analogous discrepancy and the std slightly change when the number of initial
infective varies from 10 to 100.

\begin{figure}[!h]
\centering  \includegraphics[width=80mm]{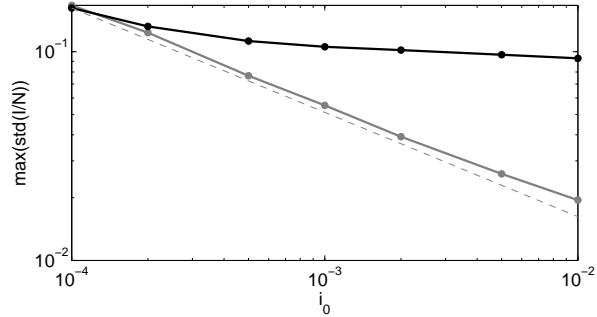}
\caption{Maximal value of the standard deviation for $I_1/N_1$ and $I_2/N_2$
process vs population $I_0$ Grey curve is for node~1, black curve is for
node~2. The dashed line has the slope corresponding to the decay law $%
N^{-1/2}$.}
\label{fig:stdImax-vs-I0}
\end{figure}

\begin{figure}[!h]
\centering  \includegraphics[width=80mm]{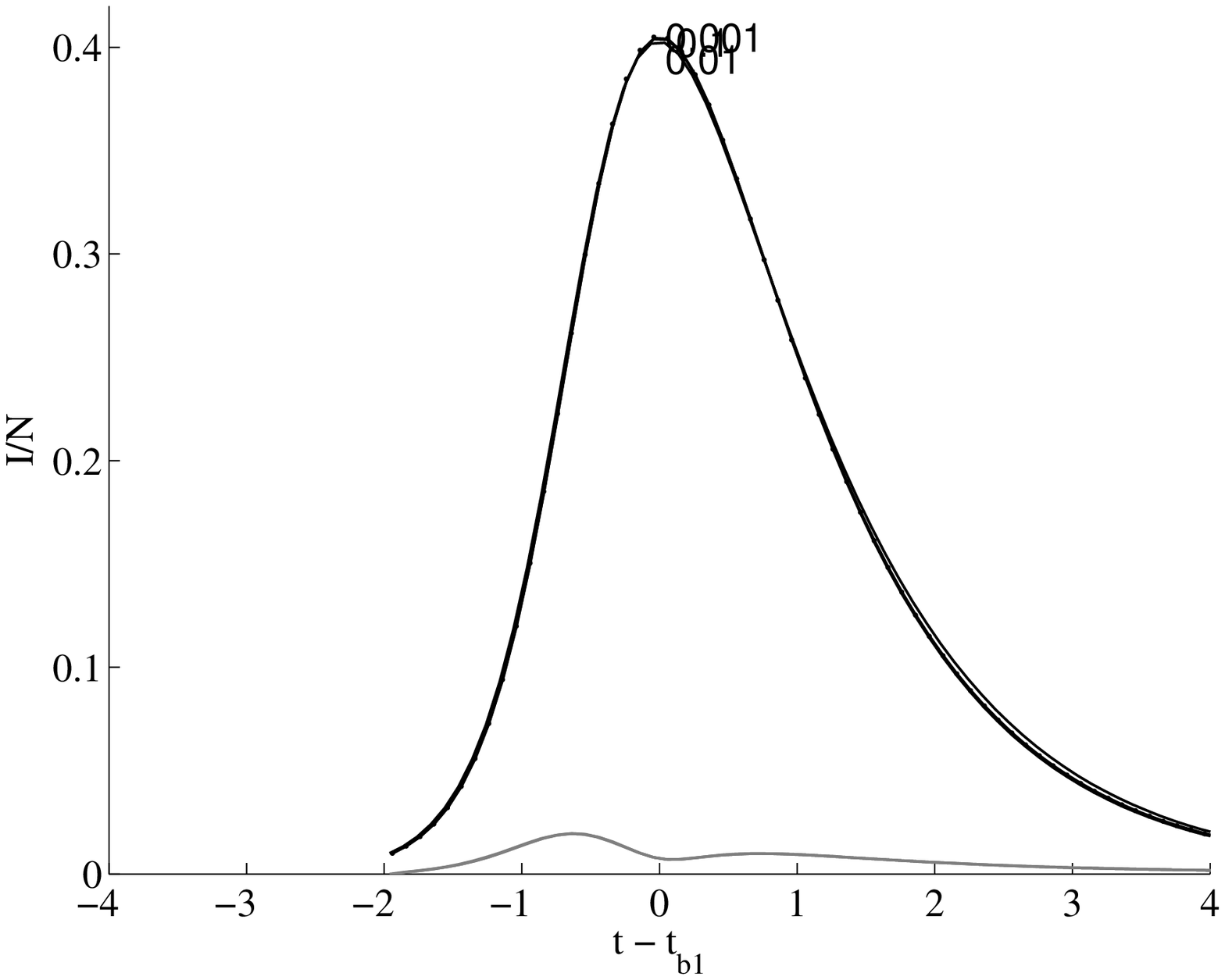} %
\includegraphics[width=80mm]{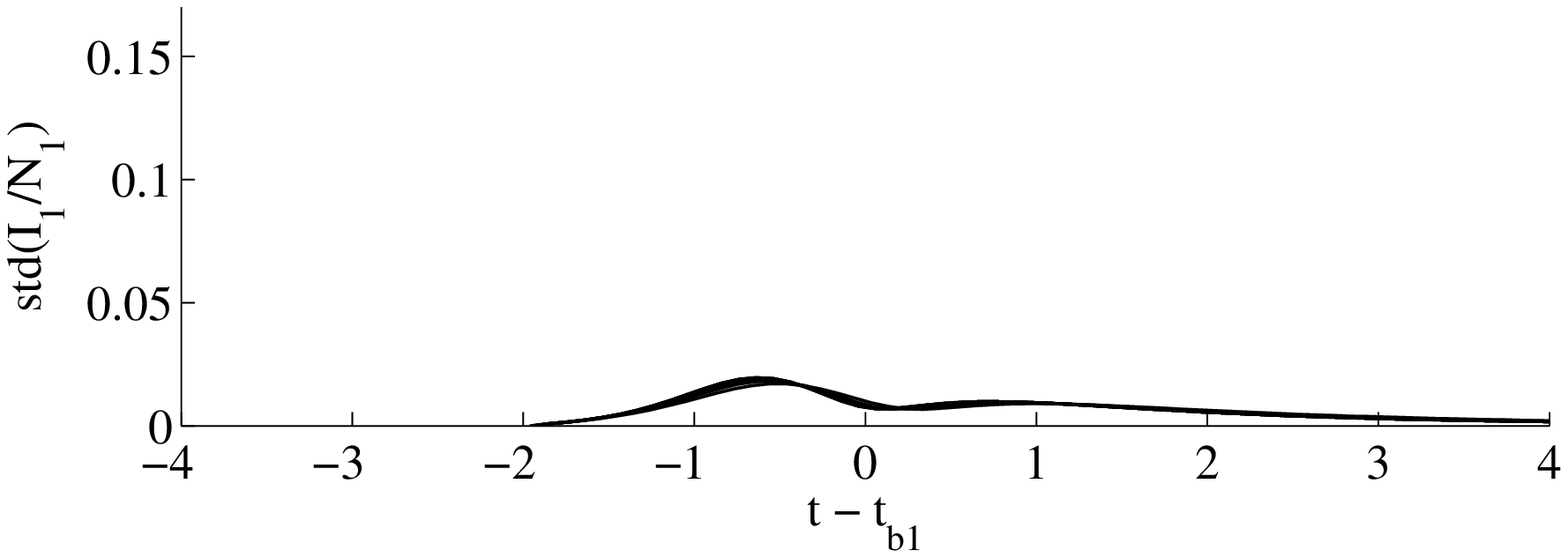}
\caption{Mean values for $I_{1}^{\Sigma }/N_{1}$ (left) and its standard
deviations $\mathrm{std}{I_{1}^{\Sigma }/N_{1}}$ (right) for different
migration coefficient $\protect\varepsilon $. Dashed lines indicate the
mean-field limit. The initial number of infectives in node~1 is indicated
near top of the corresponding curve.}
\label{fig:I1-vs-eps}
\end{figure}

\begin{figure}[!h]
\centering  \includegraphics[width=80mm]{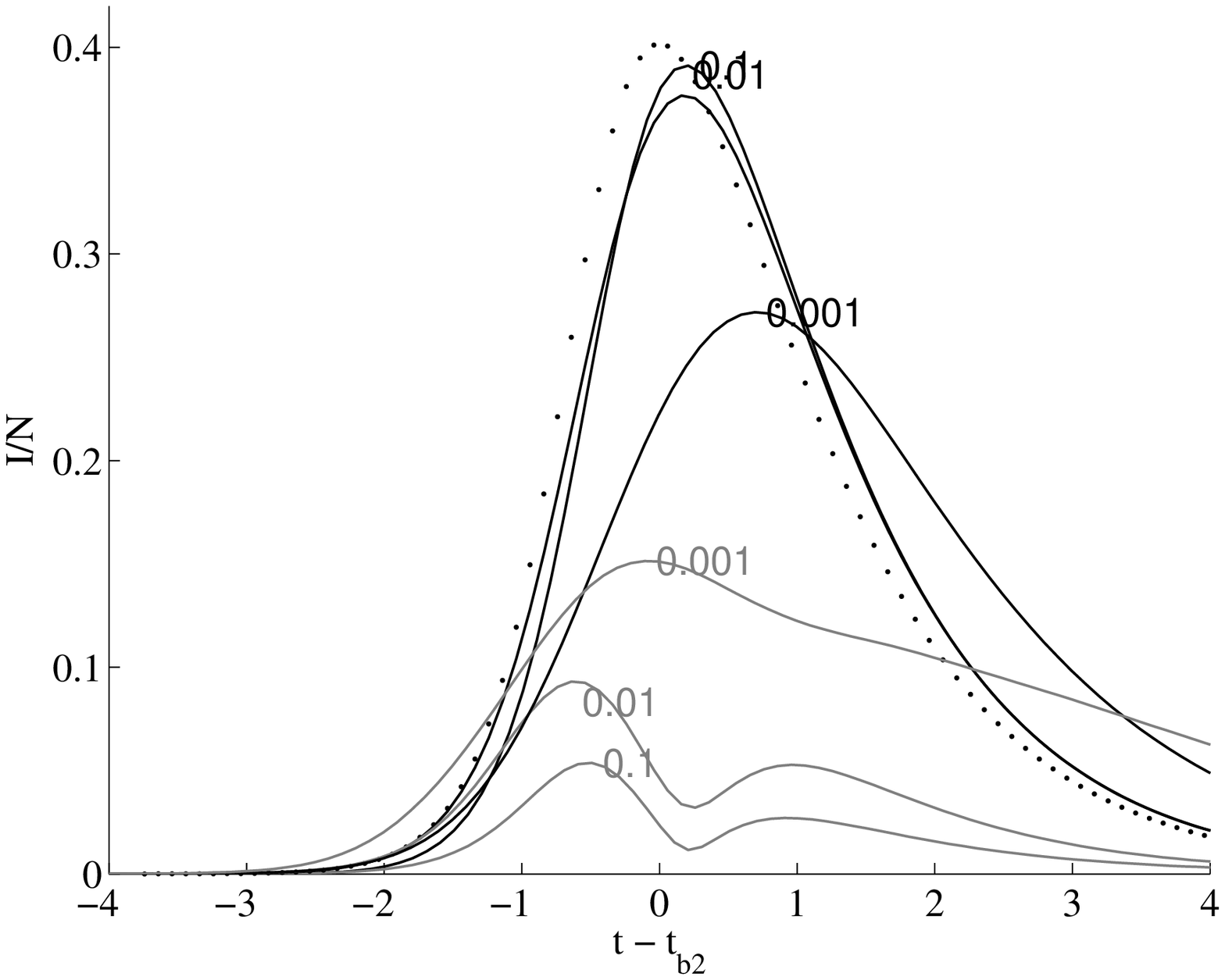} 
\includegraphics[width=80mm]{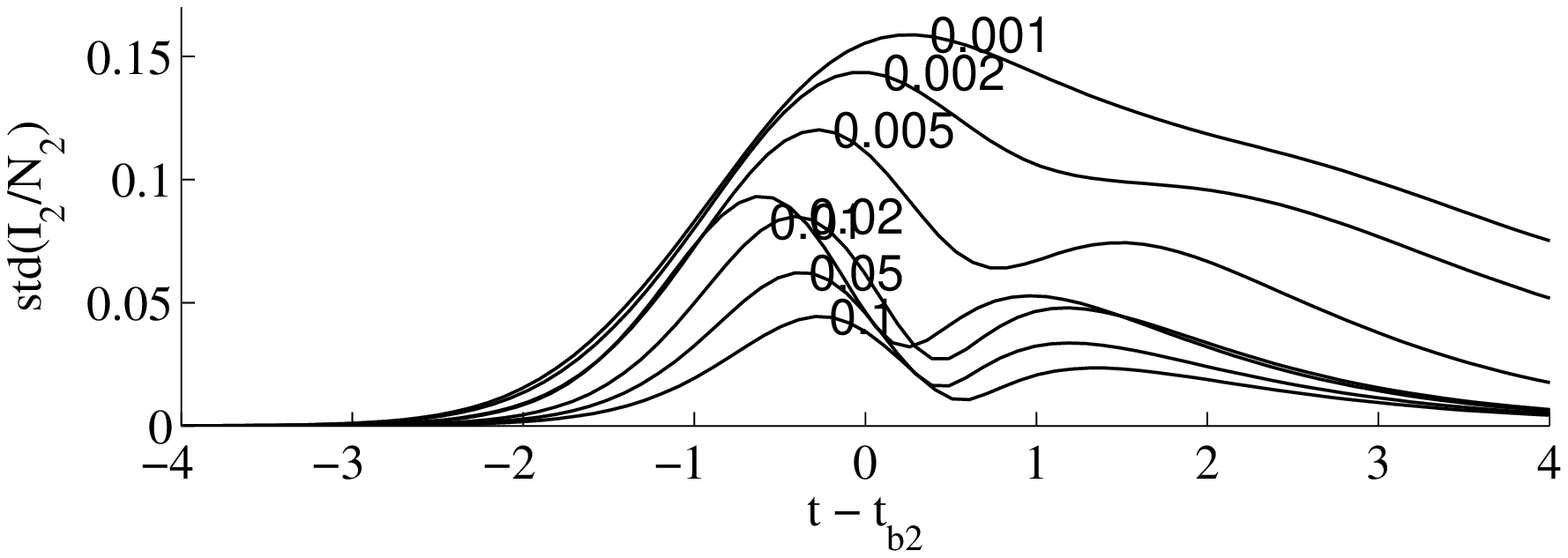}
\caption{Mean values for $I_{2}^{\Sigma }/N_{2}$ (left) and its standard
deviations $\mathrm{std}{I_{2}^{\Sigma }/N_{2}}$ (right) for different
migration coefficient $\protect\varepsilon $. Dashed lines indicate the
mean-field limit. The initial number of infectives in node~1 is indicated
near top of the corresponding curve.}
\label{fig:I2-vs-eps}
\end{figure}

In the third set of numerical experiments we study dependence of the mean number
of infectives on the coupling coefficient $\varepsilon $ (the same for all
species). It was varied in the range $10^{-4},...,10^{-1}$. The results are
plotted in Figures~\ref{fig:I1-vs-eps} and \ref{fig:I2-vs-eps}. Observe that
$\varepsilon $ practically does not affect the standard deviation of total
number of infectives in node 1. Discrepancy of the mean curve from the
mean-field curve becomes noticeable only for small $\varepsilon $: $%
\varepsilon \lesssim 0.05$.

As for the second node, the standard deviation grows monotonically with the
decrease of the coupling. That indicates the importance to account for
randomness of the epidemic process in the case of weak coupling (i.e. in the
case of relatively slow migration fluxes). 

\begin{figure}[!h]
\centering \includegraphics[width=80mm]{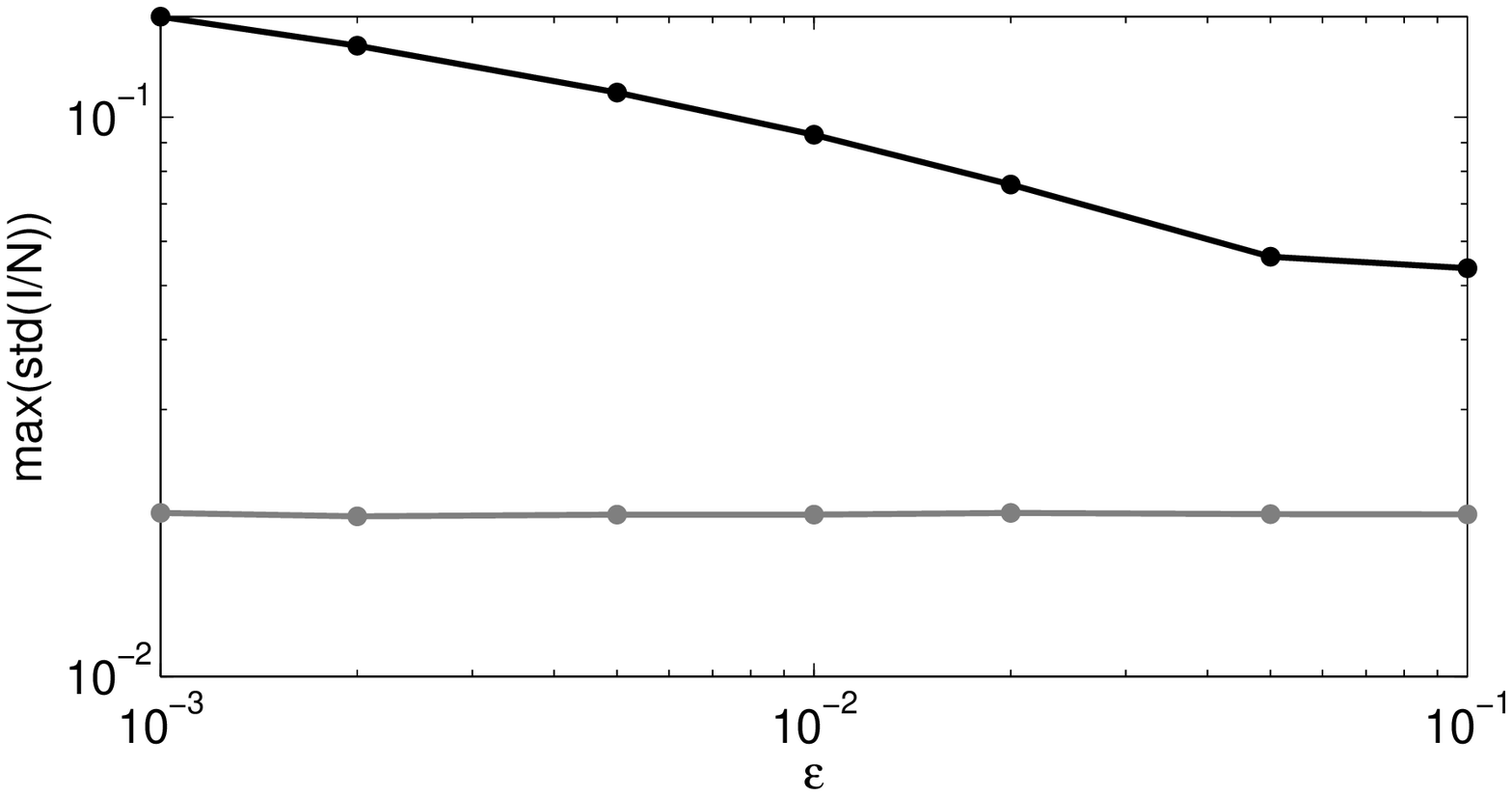} %
\includegraphics[width=80mm]{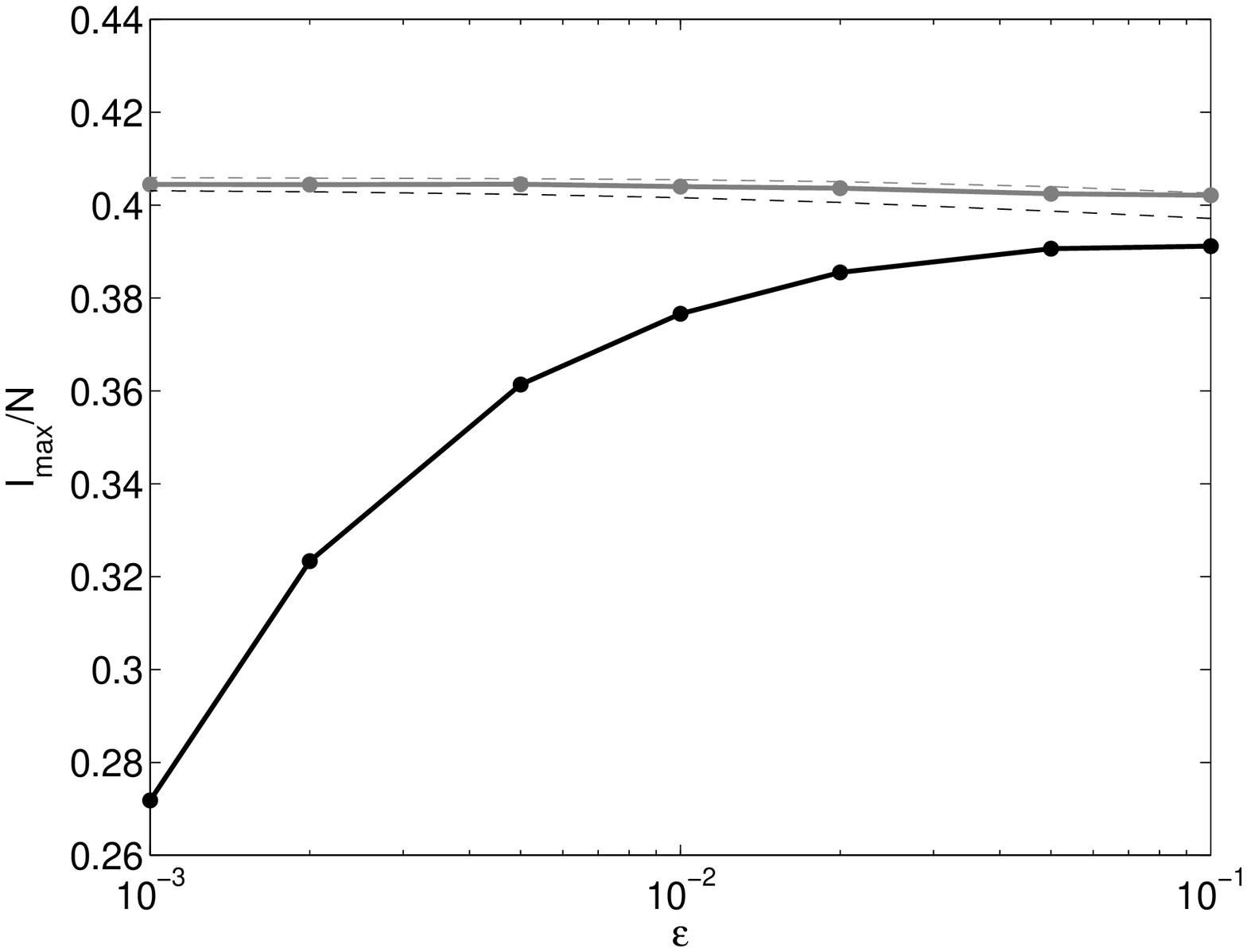}
\caption{Left: Maximal value of the standard deviation for processes $I_{1}/N_{1}$
and $I_{2}/N_{2}$ vs population $N=N_{1}=N_{2}$. Grey curve is for node~1,
black curve is for node~2. The dashed line has the slope corresponding to
the decay law $N^{-1/2}$. Right: Outbreak value for $I_{1}/N_{1}$ and $I_{2}/N_{2}$ processes vs
population $N=N_{1}=N_{2}$ Grey curve is for node~1, black curve is for
node~2. The dashed line are for the mean-field values.}
\label{fig:stdImax-vs-eps}
\end{figure}


\section{Two-stage semi-randomized model}

\label{sec:SIC}

The MC model for two coupled SIR centra can be readily generalized for an
arbitrary network of $M$ mutually interacting SIR centers:
\begin{equation}
\begin{tabular}{ll}
\hline\hline
Process & Rate \\ \hline\hline
$S_{n}\rightarrow S_{n}{-}1,I_{n}\rightarrow I_{n}{+}1$ & $\beta _{n}(I_{n}{%
+\sum_{m}}I_{mn})S_{n}$ \\
$S_{mn}\rightarrow S_{mn}{-}1,I_{mn}\rightarrow I_{mn}{+}1$ & $\beta
_{n}(I_{n}{+\sum_{m}}I_{mn})S_{mn}$ \\
$I_{n}\rightarrow I_{n}{-}1$ & $\alpha _{n}I_{n}$ \\
$I_{mn}\rightarrow I_{mn}{-}1$ & $\alpha _{n}I_{mn}$ \\ \hline
$S_{n}\rightarrow S_{n}{-}1,S_{nm}\rightarrow S_{nm}{+}1$ & $\gamma
_{nm}^{S}S_{n}$ \\
$I_{n}\rightarrow I_{n}{-}1,I_{nm}\rightarrow I_{nm}{+}1$ & $\gamma
_{nm}^{I}I_{n}$ \\
$S_{n}\rightarrow S_{n}{+}1,S_{nm}\rightarrow S_{nm}{-}1$ & $\delta
_{mn}^{S}S_{nm}$ \\
$I_{n}\rightarrow I_{n}{+}1,I_{nm}\rightarrow I_{nm}-1$ & $\delta
_{mn}^{I}I_{nm}$ \\ \hline\hline
\end{tabular}
\label{eq:MC:network}
\end{equation}%
where $m,n=1,...,M,m\neq n$.

Generically, to study migration fluxes ($\gamma _{nm}^{S,I},\delta
_{mn}^{S,I}>0$) between every pair of centers (although of different rates)
we have to consider $2M$ host and $2M(M-1)$ guest species. The correspondent
MC model will contain $4M^{2}$ fluxes. The total rate can be evaluated via $%
\nu _{\Sigma }=O(M^{2}N^{2})$. Therefore for a network containing
significant number of highly populated centra (say, main cities in the UK),
the time interval $\Delta t$ will have to be taken extremely small, and,
hence, the GPU time for a single realization will be too long to get the
necessary statistics. Thus the MC should be simplified for numerical
modelling.

\subsection{The small initial contagion (SIC) approximation}

The SIC approximation is based on the assumptions which look relevant for a
network of highly populated centers:

\begin{itemize}
\item Population in every center is high: $N_{n}\gg 1$.

\item Migration fluxes between the centers are small: $\varepsilon
_{nm}^{S,I}\ll 1$.

\item Initial number of infectives in the first contaminated center (say, $%
n=1$), is small: $I_{0}\ll N_{1}$.

\item Reproduction number exceeds unity and is not close to it in all the nodes: $\mathrm{Ro}_{n}:=\beta _{n}N_{n}/\alpha _{n}>1 + r$ where $r=O(1), r>0$. 
\end{itemize}

In these assumptions, the outbreak process in every center can be split into
the following main stages:

\begin{enumerate}
\item \textbf{Contaminating stage}: number of infectives is small $I_{n}\ll
N_{n},S_{n}\approx N_{n}$ and the fluxes of infectives caused by migration
are essential for the outbreak process (except the first node).

\item \textbf{Developed outbreak}: $I_{n}\gg 1$, when the contribution of
migration fluxes is negligible, (also the mean-field description for every
individual realization looks adequately).

\item \textbf{Recovering stage}: the node is not affected by infective
immigrants and slightly affects contamination of other nodes.
\end{enumerate}

From these assumptions follows that the outbreak dynamics in the first node
can be considered independently, and can be described by the following MC
\begin{equation}
\begin{tabular}{ll}
\hline\hline
Process & Rate \\ \hline\hline
$S_{1}\rightarrow S_{1}{-}1,I_{1}\rightarrow I_{1}{+}1$ & $\beta
_{1}S_{1}I_{1}$ \\
$I_{1}\rightarrow I_{1}{-}1$ & $\alpha _{1}I_{1}$ \\ \hline\hline
\end{tabular}
\label{MC:I1S1}
\end{equation}%
with the initial condition $I_{1}(0)=I_{0}$, $S_{1}=N_{1}-I_{0}$ studied in \cite{DG,SKG11c}. At the
contamination stage we have $S_{1}\approx N_{1}$ and the MC can be further
simplified
\begin{equation}
\begin{tabular}{ll}
\hline\hline
Process & Rate \\ \hline\hline
$I_{1}\rightarrow I_{1}{+}1$ & $\left( \beta _{1}N_{1}\right) I_{1}$ \\
$I_{1}\rightarrow I_{1}{-}1$ & $\alpha _{1}I_{1}$ \\ \hline\hline
\end{tabular}
\label{MC:I1}
\end{equation}

Epidemic dynamics in node~2 at the contamination stage ($S_{2}\approx N_{2}$%
) can be described by analogous MC with additional flux $\nu (t)$ of
infectives migrated from node~1:%
\begin{equation}
\begin{tabular}{ll}
\hline\hline
Process & Rate \\ \hline\hline
$I_{2}\rightarrow I_{2}{+}1$ & $\left( \beta _{2}N_{2}\right) I_{2}+\nu (t)$
\\
$I_{2}\rightarrow I_{2}{-}1$ & $\alpha _{2}I_{2}$ \\ \hline\hline
\end{tabular}%
\   \label{MC:I2}
\end{equation}%
with $I_{2}(0)=0$ where
\begin{equation}
\nu (t)=(\beta _{2}N_{2})I_{12}+\delta _{12}^{I}I_{21}.  \label{eq:nu}
\end{equation}

At the contamination stage processes $I_{12}(t)$ and $I_{21}(t)$ are
practically independent of process $I_{2}(t)$. Thus we have to consider MC (%
\ref{MC:I2}) with a random flux $\nu (t)$ which statistics will be specified
later.

\subsection{Calculation of moments}

First we consider a single realization of the flux $\nu (t)$ and treat it as
a deterministic function. Later we will use averaging on $\nu (t)$ to
calculate moments of distribution for $I_{2}(t)$.

\subsubsection{Calculation of PGF for a single realization $G(z,t)$.}

Let $P_{k}(t)=\mathbb{P}(I_{2}(t)=k)$ be the probability of $k$ infectives $%
I_{2}$ at instant $t$. Initial condition is
\begin{equation}
P_{0}(0)=1,\qquad P_{k>0}(0)=0  \label{eq:Pk(0)}.
\end{equation}
Kolmogorov's equations for MC (\ref{MC:I2}) are%
\begin{equation}
\textstyle{\frac{\mathrm{d}}{\mathrm{d}t}}P_{k}=\beta _{2}^{\prime
}(\left\lceil k-1\right\rceil _{0}^{N_{2}})P_{k-1}-\beta _{2}^{\prime
}kP_{k}+\displaystyle\alpha _{2}(\left\lceil k+1\right\rceil
_{0}^{N_{2}})P_{k+1}-\alpha _{2}kP_{k}+\nu P_{k-1}-\nu P_{k}.
\label{eq:nu:Pk}
\end{equation}%
Here $\beta _{2}^{\prime }=\beta _{2}N_{2}$. For the PGF $G(z,t)$
tending $N_{2}\rightarrow \infty $ (with $\beta _{2}\rightarrow 0$, $\beta
_{2}^{\prime }=const$) we obtain the following PDE
\begin{equation}
G_{t}=(z-1)\left[ \left( \beta _{2}^{\prime }z-\alpha _{2}\right) G_{z}+\nu
(t)G\right]  \label{eq:PDE-nu}
\end{equation}%
with the initial condition $G(z,0)=1$. Its solution can be written in the
integral form
\begin{equation}
G(z,t)=\exp \!\Big\{\!-\int_{0}^{t}\frac{\lambda _{2}(z{-}1)\,\nu \left(
t^{\prime }\right) \,\mathrm{d}t^{\prime }}{\beta _{2}^{\prime }(z{-}%
1)-(\beta _{2}^{\prime }z{-}\alpha _{2})\,e^{\lambda _{2}(t^{\prime }{-}t)}}%
\Big\}  \label{eq:sol-PDE-nu}
\end{equation}%
where $\lambda _{2}=\beta _{2}^{\prime }-\alpha _{2}$ is the initial growth
rate of infectives in the deterministic SIR model in the limit $%
I_{0}/N\rightarrow 0$ (limiting solution introduced in \cite{SKG08,SKG11a}).

\subsubsection{Calculation of first moment $\mathbb{E}\left[ I_{2}\left(
t\right) \right] $.}

The first conditional moment $\mu _{1}(t\,|\,\nu )=\mathbb{E}%
(I_{2}(t)\,|\,\nu )$ for fixed $\nu (t)$ is
\begin{equation}
\mu _{1}(t\,|\,\nu )=G_{z}(1,t)=\int_{0}^{t}\nu \left( t^{\prime }\right)
e^{\lambda _{2}(t-t^{\prime })}\mathrm{d}t^{\prime }  \label{eq:mu1|nu}
\end{equation}

Averaging over all realizations for $\nu (t)$ (with a time varying PDF $%
f_{\nu }(t)$): $\mu _{1}(t)=\mathbb{E}\mu _{1}(t\,|\,\nu )$
\begin{equation}
\mu _{1}(t)=\int \Big[\int_{0}^{t}f_{\nu }(t^{\prime })\nu \left( t^{\prime
}\right) e^{\lambda _{2}(t-t^{\prime })}\mathrm{d}t^{\prime }\Big]\mathrm{d}%
\nu =\int_{0}^{t}\bar{\nu}\left( t^{\prime }\right) e^{\lambda
_{2}(t-t^{\prime })}\mathrm{d}t^{\prime }  \label{eq:mu1-nu}
\end{equation}%
where $\bar{\nu}(t)=\mathbb{E}\nu (t)$. Thus the average number of
infectives in node~2 in the contamination stage relates with the
flux $\nu (t)$ via the convolution%
\begin{equation}
\mathbb{E}\left[ I_{2}\left( t\right) \right] \equiv \mu _{1}\left( t\right)
=\bar{\nu}\left( t\right) \ast e^{\lambda _{2}t}.  \label{SIC:EI2}
\end{equation}

\subsubsection{Calculation of second moment $\mathrm{var}[I_{2}(t)]$.}

To calculate it we apply the Law of Total Variation \ (e.g. \cite{Ross}):
\begin{equation}
\mathrm{var}\left[ I_{2}(t)\right] =\mathbb{E}\left[ \mathrm{var}%
(I_{2}(t)\,|\,\nu )\right] +\mathrm{var}\left[ \mathbb{E}(I_{2}(t)\,|\,\nu )%
\right].  \label{eq:var(I2)+}
\end{equation}

\textbf{1.} The first addend in (\ref{eq:var(I2)+}) can be found through the
PGF $G(z,t)$:
\begin{equation*}
\begin{array}{l}
\mathrm{var}(I_{2}(t)\,|\,\nu )=\displaystyle G_{zz}(1,t)+\mu _{1}(t\,|\,\nu
)-\mu _{1}^{2}(t\,|\,\nu ) \\
\displaystyle=\frac{2\beta _{2}^{\prime }}{\lambda _{2}}\int_{0}^{t}\nu
(t^{\prime })\left[ e^{2\lambda _{2}(t-t^{\prime })}-e^{\lambda
_{2}(t-t^{\prime })}\right] \mathrm{d}t^{\prime }+\mu _{1}(t\,|\,\nu ).%
\end{array}%
\end{equation*}%
After the averaging through $\nu $ we obtain
\begin{equation*}
\mathbb{E}\left[ \mathrm{var}(I_{2}(t)\,|\,\nu )\right] =\displaystyle\mu
_{1}(t)+\frac{2\beta _{2}^{\prime }}{\lambda _{2}}\int_{0}^{t}\bar{\nu}%
(t^{\prime })\left[ e^{2\lambda _{2}(t-t^{\prime })}-e^{\lambda
_{2}(t-t^{\prime })}\right] \mathrm{d}t^{\prime }.
\end{equation*}%
Thus a first addend in (\ref{eq:var(I2)+}) can written as a sum of two
convolutions
\begin{equation}
\mathbb{E}\left[ \mathrm{var}(I_{2}(t)\,|\,\nu )\right] =\frac{2\beta
_{2}^{\prime }}{\lambda _{2}}\bar{\nu}(t)\ast e^{2\lambda _{2}t}+(1-\frac{%
2\beta _{2}^{\prime }}{\lambda _{2}})\bar{\nu}(t)\ast e^{\lambda _{2}t}
\label{eq:var(I2):1}
\end{equation}%
where
$\bar{\nu}(t)=\beta _{2}^{\prime }\bar{I}_{12}(t)+\delta _{12}^{I}\bar{I}%
_{21}(t)$.

\textbf{2.} To calculate the second addend in (\ref{eq:var(I2)+}), we
temporary add $\mu _{1}^{2}$ to it. Now it can be expressed via the
covariance of flux $\nu (t)$:%
\begin{equation}
\begin{array}{l}
\displaystyle\mathrm{var}\left[ \mathbb{E}(I_{2}(t)\,|\,\nu )\right] +\mu
_{1}^{2}(t)=\mathbb{E}\left( \int_{0}^{t}\nu \left( t^{\prime }\right)
e^{\lambda _{2}(t-t^{\prime })}\mathrm{d}t^{\prime }\right) ^{2} \\[0.5pc]
\displaystyle=\int_{0}^{t}\int_{0}^{t}\mathbb{E}\left[ \nu \left( t^{\prime
}\right) \nu \left( t^{\prime \prime }\right) \right] e^{\lambda
_{2}(t-t^{\prime })}\mathrm{d}t^{\prime }e^{\lambda _{2}(t-t^{\prime \prime
})}\mathrm{d}t^{\prime \prime }.%
\end{array}
\label{eq:var(I2)a}
\end{equation}%
The function in the integrand can be represented as a sum
\begin{equation}
\mathbb{E}\left[ \nu \left( t^{\prime }\right) \nu \left( t^{\prime \prime
}\right) \right] =\mathrm{cov}\left[ \nu \left( t^{\prime }\right) ,\nu
\left( t^{\prime \prime }\right) \right] +\bar{\nu}(t^{\prime })\bar{\nu}%
(t^{\prime \prime }).  \label{eq:E[nu(t')*nu(t")]}
\end{equation}%
Integration of the second addend in (\ref{eq:E[nu(t')*nu(t")]}) gives just
the temporary added term%
\begin{equation*}
\int_{0}^{t}\int_{0}^{t}\bar{\nu}(t^{\prime })\bar{\nu}(t^{\prime \prime
})e^{\lambda _{2}(t-t^{\prime })}\mathrm{d}t^{\prime }e^{\lambda
_{2}(t-t^{\prime \prime })}\mathrm{d}t^{\prime \prime }=\mu _{1}^{2}(t).
\end{equation*}%
Thus the second addend in (\ref{eq:var(I2)+}) can be written through the
following integral%
\begin{equation}
\mathrm{var}\left[ \mathbb{E}\big(I_{2}(t)\;|\;\nu \big)\right]
=\int_{0}^{t}\int_{0}^{t}\mathrm{cov}\left[ \nu \left( t^{\prime }\right)
,\nu \left( t^{\prime \prime }\right) \right] e^{\lambda _{2}(t-t^{\prime })}%
\mathrm{d}t^{\prime }e^{\lambda _{2}(t-t^{\prime \prime })}\mathrm{d}%
t^{\prime \prime }.  \label{eq:var(I2)b}
\end{equation}%
in which we have to calculate the covariance of flux $\nu (t)$.

If flux $\nu (t)$ is a random process controlled by a MC, calculation of its
covariance is a complicated task, and consideration of this needs a separate
work. Remind that the flux is a linear combination of two MC processes (\ref{eq:nu}): $\nu
(t)=\beta _{2}^{\prime }I_{12}(t)+\delta _{12}^{I}I_{21}(t)$. Here we
approximate $I_{12}(t)$ and $I_{21}(t)$ by two mutually independent Poisson processes
with variable rates $\frac{\mathrm{d}}{\mathrm{d}t}\bar{I}_{12}(t)$ and $%
\frac{\mathrm{d}}{\mathrm{d}t}\bar{I}_{21}(t)$, respectively, where $\bar{I}%
_{12}(t)$ and $\bar{I}_{21}(t)$ are calculated below. In this approximation,
using the independence of increments of the inhomogeneous Poisson flow (e.g.
\cite{SK}) we can write%
\begin{eqnarray*}
\mathrm{cov}\left[ I_{12}\left( t^{\prime }\right) ,I_{12}\left( t^{\prime
\prime }\right) \right] &\approx &\bar{I}_{12}(\min \left\{ t^{\prime
},t^{\prime \prime }\right\} ), \\
\mathrm{cov}\left[ I_{21}\left( t^{\prime }\right) ,I_{21}\left( t^{\prime
\prime }\right) \right] &\approx &\bar{I}_{21}(\min \left\{ t^{\prime
},t^{\prime \prime }\right\} ), \\
\mathrm{cov}\left[ I_{12}\left( t^{\prime }\right) ,I_{21}\left( t^{\prime
\prime }\right) \right] &\approx &0.
\end{eqnarray*}

We justify this approximation numerically below. Thus, for the covariance of
the flux we have%
\begin{equation}
\mathrm{cov}\left[ \nu \left( t^{\prime }\right) ,\nu \left( t^{\prime
\prime }\right) \right] =\varpi (\min \left\{ t^{\prime },t^{\prime \prime
}\right\} ),\quad \varpi (t)\equiv \left( \beta _{2}^{\prime }\right) ^{2}%
\bar{I}_{12}(t)+\left( \delta _{12}^{I}\right) ^{2}\bar{I}_{21}(t).
\label{eq:cov:Po}
\end{equation}%
Equation (\ref{eq:cov:Po}) holds true because the second central moment of the
Poisson process coincides with the first moment. In this approximation, it is
sufficient to compute the first moment of flux $\nu (t)$ in order to compute
the second moment for $I_{2}$.

With the account for (\ref{eq:cov:Po}), we split integral (\ref{eq:var(I2)b}%
) into two parts:%
\begin{eqnarray*}
\mathrm{var}\left[ \mathbb{E}\big(I_{2}(t)\;|\;\nu \big)\right] &\approx
&J_{1}+J_{2} \\
J_{1} &=&\int_{0}^{t}\int_{0}^{t^{\prime }}\varpi (t^{\prime \prime
})e^{\lambda _{2}(t-t^{\prime \prime })}\mathrm{d}t^{\prime \prime
}e^{\lambda _{2}(t-t^{\prime })}\mathrm{d}t^{\prime } \\
J_{2} &=&\int_{0}^{t}\int_{t^{\prime }}^{t}\varpi (t^{\prime })e^{\lambda
_{2}(t-t^{\prime \prime })}\mathrm{d}t^{\prime \prime }e^{\lambda
_{2}(t-t^{\prime })}\mathrm{d}t^{\prime }.
\end{eqnarray*}
Integrating $J_{1}$ by parts and $J_{2}$ directly we obtain that they both
give the same answer%
\begin{equation*}
J_{1}=J_{2}=\frac{1}{\lambda _{2}}\varpi (t)\ast e^{2\lambda _{2}t}-\frac{1}{%
\lambda _{2}}\varpi (t)\ast e^{\lambda _{2}t}.
\end{equation*}
Finally combining the above results we have%
\begin{equation}
\begin{array}{l}
\mathrm{var}(I_{2})=\displaystyle\frac{4\left( \beta _{2}^{\prime }\right)
^{2}}{\lambda _{2}}\bar{I}_{12}\ast e^{2\lambda _{2}t}+\Big[ \frac{2\beta
_{2}^{\prime }\delta _{12}^{I}}{\lambda _{2}}+\frac{2\left( \delta
_{12}^{I}\right) ^{2}}{\lambda _{2}}\Big] \bar{I}_{21}\ast e^{2\lambda _{2}t}
\\[0.7pc]
\displaystyle\;{}-\Big[ \frac{4\left( \beta _{2}^{\prime }\right) ^{2}}{\lambda
_{2}}-\beta _{2}^{\prime }\Big] \bar{I}_{12}\ast e^{\lambda _{2}t}-\Big[
\frac{2\beta _{2}^{\prime }\delta _{12}^{I}}{\lambda _{2}}+\frac{2\left(
\delta _{12}^{I}\right) ^{2}}{\lambda _{2}}-\delta _{12}^{I}\Big] \bar{I}%
_{21}\ast e^{\lambda _{2}t}.%
\end{array}
\label{eq:var(I2)}
\end{equation}

\subsubsection{Computation of the average flux $\bar{\protect\nu}(t)$.}

It is natural to split the total flux into two parts $\nu (t)=\beta
_{2}^{\prime }I_{12}+\delta _{12}^{I}I_{21}=\nu _{12}(t)+\nu _{21}(t)$. Flux process $\nu _{12}(t)$ described by the MC
\begin{equation}
\begin{tabular}{ll}
\hline\hline
Process & Rate \\ \hline\hline
$I_{12}\rightarrow I_{12}{+}1$ & $\gamma _{12}^{I}I_{1}$ \\
$I_{12}\rightarrow I_{12}{-}1$ & $\left( \delta _{21}^{I}+\alpha _{2}\right)
I_{12}$ \\ \hline\hline
\end{tabular}
\label{MC:I12}
\end{equation}%
(with $I_{12}(0)=0$) coincides with that described by (\ref{MC:I2}) if we set $\beta
_{2}^{\prime }\leftarrow 0$, $\nu (t)\leftarrow \gamma _{12}^{I}I_{1}$, $%
\alpha _{2}\leftarrow \left( \delta _{21}^{I}+\alpha _{2}\right) $. Then we
can immediately write down a solution for the PGF
\begin{equation*}
G(z,t)=\exp \!\big\{\gamma _{12}^{I}\left( z{-}1\right)
\!\!\!\int_{0}^{t}I_{1}\left( t^{\prime }\right) e^{-\left( \delta
_{21}^{I}+\alpha _{2}\right) (t^{\prime }-t)}\mathrm{d}t^{\prime }\!\big\}
\end{equation*}%
and the first moment%
\begin{equation}
\bar{I}_{12}=\gamma _{12}^{I}\int_{0}^{t}\bar{I}_{1}\left( t^{\prime
}\right) e^{-(\delta _{21}^{I}+\alpha _{2})(t-t^{\prime })}\mathrm{d}%
t^{\prime }.  \label{eq:EI12}
\end{equation}%

Flux process $\nu _{21}$ is more complicated and can be described by the
following MC%
\begin{equation}
\begin{tabular}{ll}
\hline\hline
Process & Rate \\ \hline\hline
$S_{21}\rightarrow S_{21}+1$ & $\gamma _{21}^{S}N_{2}$ \\
$S_{21}\rightarrow S_{21}-1$ & $\delta _{12}^{S}S_{21}$ \\
$I_{21}\rightarrow I_{21}+1,S_{21}\rightarrow S_{21}-1$ & $\beta
_{1}I_{1}S_{21}$ \\
$I_{21}\rightarrow I_{21}-1$ & $\delta _{12}^{I}I_{21}$ \\ \hline\hline
\end{tabular}
\label{MC:nu21}
\end{equation}%
with the initial conditions $S_{21}(0)=\varepsilon _{21}^{S}N_{2}$, $%
I_{21}(0)=0$. If we split the third event into two independent events $%
I_{21}\rightarrow I_{21}+1$ and $S_{21}\rightarrow S_{21}-1$ with the same
rate, we can split MC (\ref{MC:nu21}) into two MCs. The first MC describes
migration of host susceptives from node~2 to node~1 and their possible
removal due to contamination:%
\begin{equation}
\begin{tabular}{ll}
\hline\hline
Process & Rate \\ \hline\hline
$S_{21}\rightarrow S_{21}+1$ & $\gamma _{21}^{S}N_{2}$ \\
$S_{21}\rightarrow S_{21}-1$ & $\left( \delta _{12}^{S}+\beta
_{1}I_{1}\right) S_{21}$ \\ \hline\hline
\end{tabular}
\label{MC:S21}
\end{equation}%
It is independent of the second MC with the rates%
\begin{equation}
\begin{tabular}{ll}
\hline\hline
Process & Rate \\ \hline\hline
$I_{21}\rightarrow I_{21}+1$ & $\beta _{1}I_{1}S_{21}$ \\
$I_{21}\rightarrow I_{21}-1$ & $\left( \delta _{12}^{I}+\alpha _{1}\right)
I_{21}$ \\ \hline\hline
\end{tabular}
\label{MC:I21}
\end{equation}%
which describes\ migration of susceptibles to a neighbor node, their
contamination there and return to the host node as infected species.

First, we study the first MC: (\ref{MC:S21}). In accordance with \ (\ref%
{eq:S0-distribution}) it has the binomial initial distribution:%
\begin{equation}
P_{k}(0)=\binom{N_{2}}{k}\left( \varepsilon _{21}^{S}\right) ^{k}\left(
1-\varepsilon _{21}^{S}\right) ^{N_{2}-k}.  \label{SIC:S210}
\end{equation}%
The probabilities $P_{k}(t)=\mathbb{P}(S_{21}(t)=k)$ of $k$ guest
susceptibles $S_{21}$ at instant $t$ satisfy Kolmogorov's equations for MC (%
\ref{MC:I2})
\begin{equation}
\textstyle{\frac{\mathrm{d}}{\mathrm{d}t}}P_{k} = \nu \left(
P_{k-1}-P_{k}\right) +\alpha \left[ (\left\lceil k+1\right\rceil
_{0}^{N_{2}})P_{k+1}-kP_{k}\right] .  \label{eq:S21:Pk}
\end{equation}%
where $\nu =\gamma _{21}^{S}N_{2}$,\ $\alpha =\left( \delta _{12}^{S}+\beta
_{1}I_{1}\right) $. System (\ref{eq:S21:Pk}) implies the following PDE for
MGF $G(z,t)=\sum_{k=0}^{\infty }z^{k}P_{k}(t)$
\begin{equation}
G_{t}=(z-1)\left[ -\alpha (t)G_{z}+\nu G\right].  \label{eq:S21:PDE}
\end{equation}%
The initial condition is
\begin{equation}
G(z,0)=\sum\limits_{k=0}^{N}z^{k}\binom{N}{k}\varepsilon ^{k}(1-\varepsilon
)^{N-k}=(1-\varepsilon +\varepsilon z)^{N}.  \label{SIC:G0}
\end{equation}%
The initial value problem (\ref{eq:S21:PDE})--(\ref{SIC:G0}) admits the
explicit solution
\begin{equation}
G(z,t)=\left[ 1+\varepsilon (z-1)\phi (t)\right] ^{N}\exp \left\{ \nu
(z-1)\phi (t)\int_{0}^{t}\frac{\mathrm{d}t^{\prime }}{\phi (t^{\prime })}%
\right\}.  \label{SIC:G}
\end{equation}%
where $\phi (t)=\exp \left\{ -\int_{0}^{t}\alpha (t^{\prime })\mathrm{d}%
t^{\prime }\right\} $. From here we have%
\begin{equation}
\displaystyle\mathbb{E}S_{21}(t)=G_{z}(1,t)= N_{2}\left[ \varepsilon
_{21}^{S}+\gamma _{21}^{S}\int_{0}^{t}\mathrm{d}t^{\prime }/\phi (t^{\prime
})\right] \phi (t).  \label{eq:ES21}
\end{equation}

Now for process (\ref{MC:I21}) in analogy with processes (\ref{MC:I2}) and (%
\ref{MC:I12}) we can immediately write%
\begin{equation}
\bar{I}_{21}=\beta _{1}\int_{0}^{t}e^{-\left( \delta _{12}^{I}+\alpha
_{2}\right) (t-t^{\prime })}\mathbb{E}\left[ I_{1}\left( t^{\prime }\right)
S_{21}(t^{\prime })\right] \mathrm{d}t^{\prime }.  \label{eq:EI21}
\end{equation}%
Neglecting the mutual dependence of processes $S_{21}(t)$ and $I_{1}(t)$ we
approximate$\ $%
\begin{equation}
\mathbb{E}\left[ S_{21}(t)I_{1}(t)\right] \approx \mathbb{E}S_{21}(t)\bar{I}%
_{1}(t).  \label{eq:E(S21*I1)}
\end{equation}%
Below we show numerically that it is satisfactory approximation for our
applications.

\subsection{The second stage}

Remind that equations (\ref{eq:mu1-nu}), (\ref{eq:var(I2)b}), (\ref%
{eq:cov:Po}), (\ref{eq:var(I2)}), (\ref{eq:EI12}), (\ref{eq:EI21}), (\ref%
{eq:E(S21*I1)}) are valid at the contamination stage only ($S_{2}\approx
N_{2}$). They allow us to calculate the first and second moments for the
number of infectives without modelling the random process directly. To
evaluate the moments at the developed outbreak we use the same approach as
in \cite{SKG11c}: by approximating the outbreak via the mean field solution
for a single SIR node with random initial conditions.

For this aim we define the intermediate time $t_{\ast }$ such that the
number of infective is large enough to use the mean field solution but the
number of infective still slightly deviate from $N_{2}$: $\hat{I}%
_{2}(t_{\ast })\gg 1$ and $N_{2}-\hat{S}_{2}\ll N_{2}$.

Then we generate $L$\ times a random number $X$ lognormally distributed (to
guarantee the positiveness) with mean $\bar{I}_{2}(t_{\ast })$ and variance $%
\mathrm{var}(I_{2}(t_{\ast }))$ and integrate the classical SIR equations: $%
\frac{\mathrm{d}}{\mathrm{d}t}\hat{S}_{2}=-\beta _{2}\hat{S}_{2}\hat{I}_{2}$%
, $\frac{\mathrm{d}}{\mathrm{d}t}\hat{I}_{2}=(\beta _{2}\hat{S}_{2}-\alpha
_{2})\hat{I}_{2}$ with initial condition $\hat{I}_{2}(t_{\ast })=X$,\
\begin{equation*}
\hat{S}_{2}(t_{\ast })=-W_{-1}\left[ -\beta _{2}^{\prime }N_{2}\exp \left\{
\beta _{2}^{\prime }(X-N_{2})\right\} \right] /\beta _{2}^{\prime }
\end{equation*}%
where $W_{k}[\cdot ]$ is the $k$th branch of the Lambert function \cite%
{Lambert}.

Let us emphasize that in the classical SIR model, the numbers of susceptives
and infectives are related as $I=N-S+\ln \left[ S/(N-I_{0})\right] /\beta
^{\prime }$ where $\beta ^{\prime }=\beta N$ (cf. \cite{DG}). Resolving this
relation with respect to $S$ we can write
\begin{equation*}
S=-W_{k}\left[ -\beta ^{\prime }(N-I_{0})\exp \left\{ \beta ^{\prime
}(I-N)\right\} \right] /\beta ^{\prime }
\end{equation*}%
where $k=-1$ for the growing part and $k=0$ for the decaying part of the
outbreak. If the outbreak is triggered by a infinitesimal number of
infectives we can set $S=-W_{k}\left[ -\beta ^{\prime }N\exp \left\{ \beta
^{\prime }(I-N)\right\} \right] /\beta ^{\prime }$.

Also note that it is natural to approximate the solution to a standard SIR
model by the limiting solution ($I_{0}/N\rightarrow 0$) introduced in \cite%
{SKG08}. The limiting solution is independent of the initial condition,
therefore it is not necessity to integrate the ODEs $L$ times but only once.

Thus the proposed two-stage model of a coupled randomized epidemic centers
allows us to calculate its first moments much faster than the direct
simulation summarized in Table~\ref{tab:MC}.

\subsection{Comparison with numerical simulation}

To show the accuracy of the proposed model we compare the solution obtained
by different approaches. Again the basic model of Section~\ref{sec:numerics}
is used: $N_{1}=N_{2}=10$k, $\beta _{1,2}^{\prime }=4$, $\alpha _{1,2}=1$%
, $\varepsilon _{1,2}^{S,I}=0.01$, $\tau _{1,2}^{S,I}=5$, $I_{0}/N_{1}=0.01$%
 but also model with the smaller population $N_{1}=N_{2}=2$k. We compare (i) the full randomized model described in Table~\ref{tab:MC}
which we regard as a benchmark; (ii) the SIC approximate randomized model
where only flux of infectives from node~1 to node~2 is accounted, it is
described in Table~\ref{tab:MC:SIC}; (iii) the two-stage semi-randomized
model proposed in this section above. 

In the two-stage model we take time $t_{\ast }=2.0$ for transition from a
contamination randomized stage to the mean-field stage with random initial
condition. The expected number of infectives in node~2 at time~$t=2.0$ is $%
100$ which is large enough and at the same time much smaller than the node
population. We take $L=10^4$ for number of realization in the second stage
to evaluate the moments.

In the SIC approximation the randomized model presented in Table~\ref%
{tab:MC:SIC} comprises four consequently independent MCs. Processes 1 and 2
represent independent outbreak in node~1 (\ref{MC:I1S1}). Processes 3 and 4
represent migration of its infectives to node~2 (\ref{MC:I12}). Processes 5
and 6 represent migration of host susceptives from node 2 to node 1 and
their possible removal due to contamination; they are analogous to MC (\ref%
{MC:S21}) with $N_{2}$ substituted by $S_{2}$ to be valid for all the
stages. Analogously processes 7 and 8 represent contamination of $S_{21}$ in
the first node and migration of appeared infectives $I_{21}$ to their host
node (\ref{MC:I21}). Finally processes 9--10 represent the outbreak in
node~2 ; they are analogous to MC with an additional flux (\ref{MC:I2}) with
the accurate account of the number of succeptives $S_{2}$ at all the stages.

\begin{table}[tbp]
\caption{Markov's chain for two coupled SIR nodes in the SIC approximation}
\label{tab:MC:SIC}\centering
\begin{tabular}{lll}
\hline\hline
\# & Process & Rate \\ \hline\hline
1 & $%
\begin{array}{l}
S_{1}\rightarrow S_{1}{-}1 \\
I_{1}\rightarrow I_{1}{+}1%
\end{array}%
$ & $\beta _{1}I_{1}S_{1}$ \\
2 & $I_{1}\rightarrow I_{1}{-}1$ & $\alpha _{1}I_{1}$
\rule[-0.5pc]{0pc}{1.5pc} \\ \hline
3 & $I_{12}\rightarrow I_{12}{+}1$ & $\gamma _{12}^{I}I_{1}$ \\
4 & $I_{12}\rightarrow I_{12}-1$ & $\delta _{21}^{I}I_{12}$ \\ \hline
5 & $S_{21}\rightarrow S_{21}{+}1$ & $\gamma _{21}^{S}S_{2}$ \\
6 & $S_{21}\rightarrow S_{21}{-}1$ & $\left( \delta _{12}^{S}+\beta
_{1}I_{1}\right) S_{21}$ \\ \hline
7 & $I_{21}\rightarrow I_{21}{+}1$ & $\beta _{1}I_{1}S_{21}$ \\
8 & $I_{21}\rightarrow I_{21}{-}1$ & $\alpha _{1}I_{21}$ \\ \hline
9 & $%
\begin{array}{l}
S_{2}\rightarrow S_{2}{-}1 \\
I_{2}\rightarrow I_{2}{+}1%
\end{array}%
$ & $\beta _{2}(I_{2}+I_{12})S_{2}+\delta _{12}^{I}I_{21}$ \\
10 & $I_{2}\rightarrow I_{2}{-}1$ & $\alpha _{2}I_{2}$
\rule[-0.5pc]{0pc}{1.5pc} \\ \hline\hline
\end{tabular}%
\end{table}

\begin{figure}[h]
\centering \includegraphics[width=65mm]{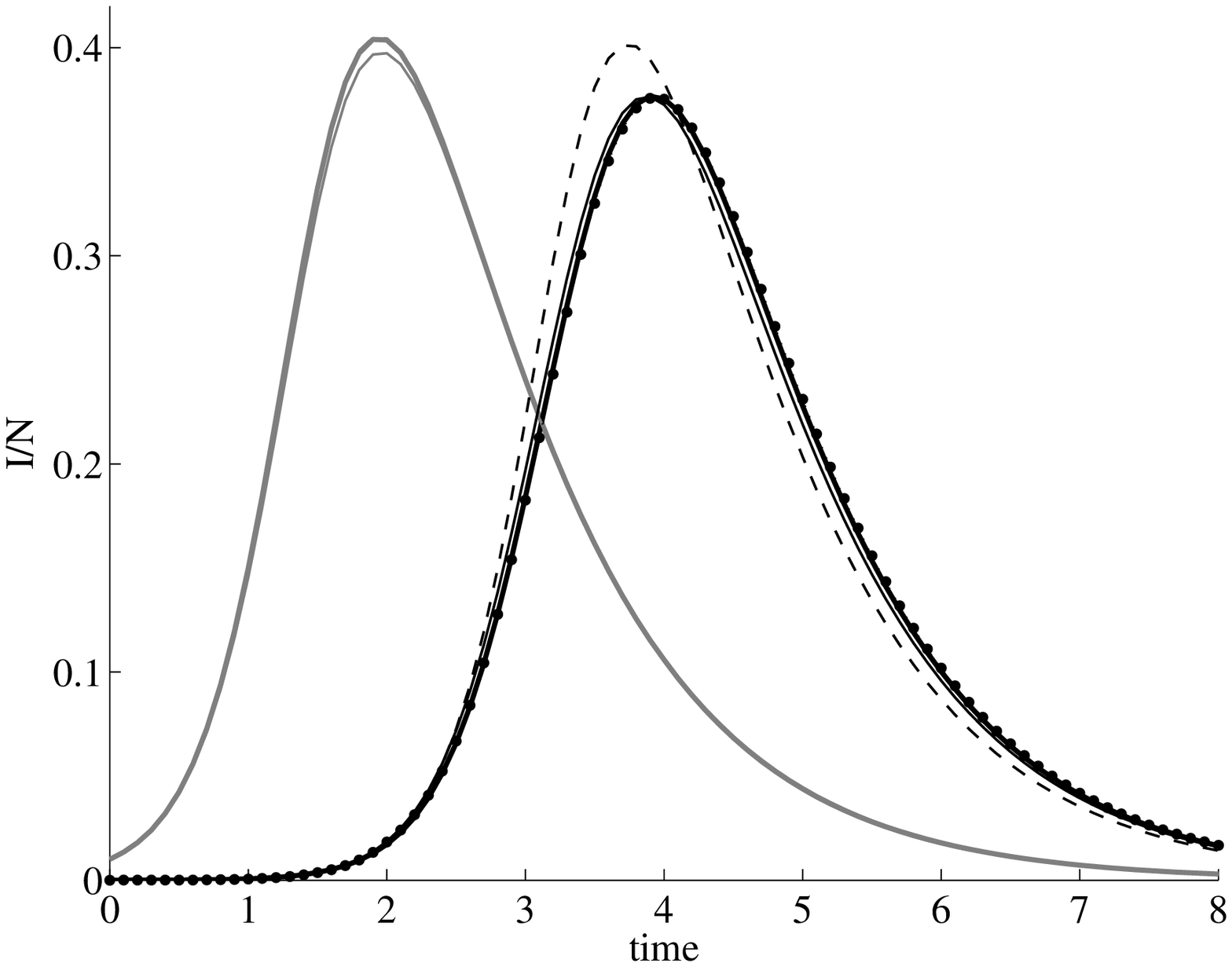} %
\includegraphics[width=65mm]{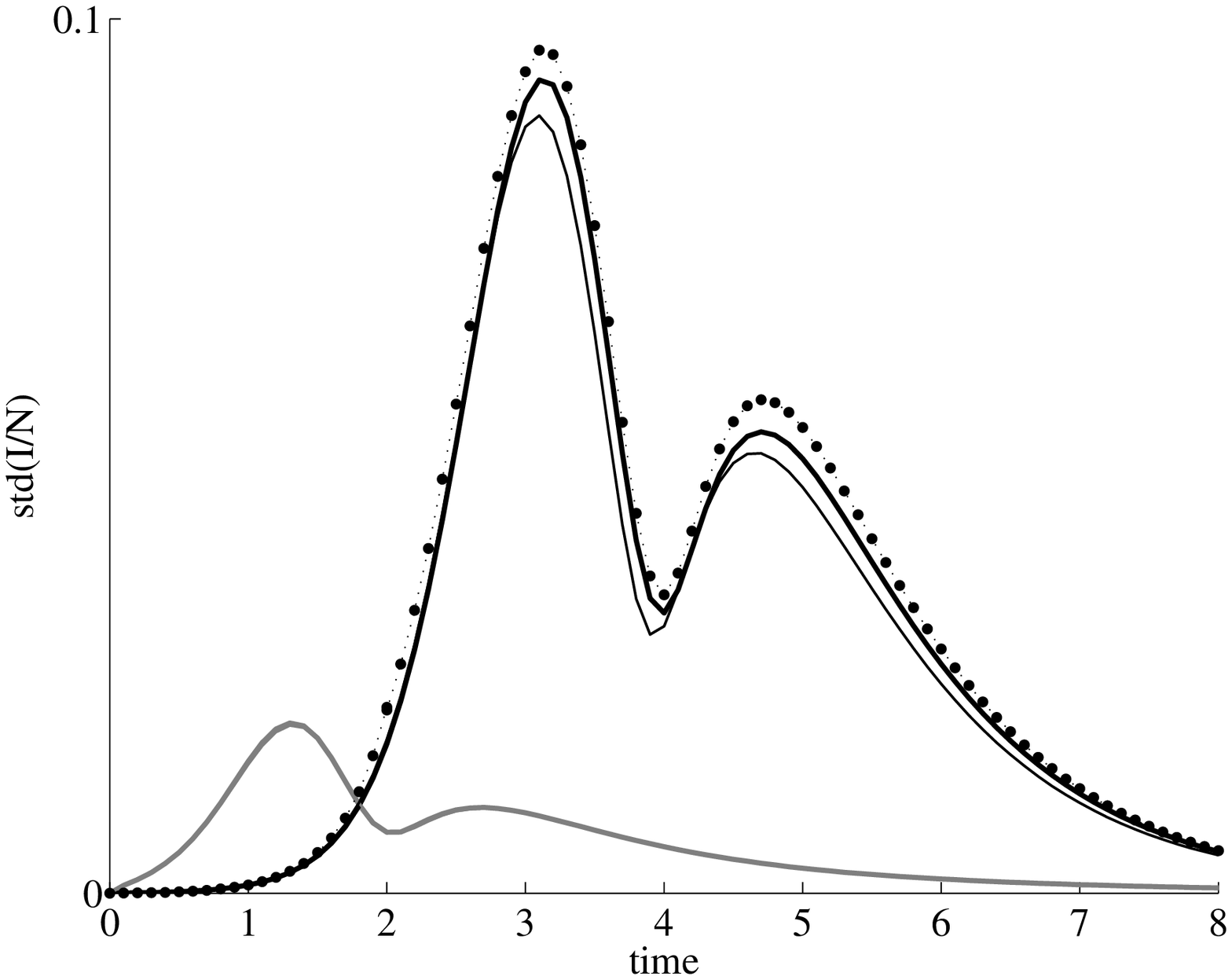}
\caption{Comparison of different approximation for the mean value of
infectives (left) and its standard deviation (right) for the populations $N_1=N_2=10$k.
Bold line -- full
randomized model (Table~1), thin line -- approximate randomized model
(Table~2), line with dots -- the proposed two-stage semi-randomized model,
dashed line -- the mean field solution.}
\label{fig:SIC10k}
\end{figure}

The results of computation are presented in Figures~\ref{fig:SIC10k} and \ref%
{fig:SIC2k}. Here the bold lines indicate the full randomized model, the
thin solid lines indicate the SIC approximated model, the line with dots
indicate the two-stage semi-randomized model. Also the mean-field solution
is presented as well and indicated by dashed lines.

\begin{figure}[h]
\centering \includegraphics[width=60mm]{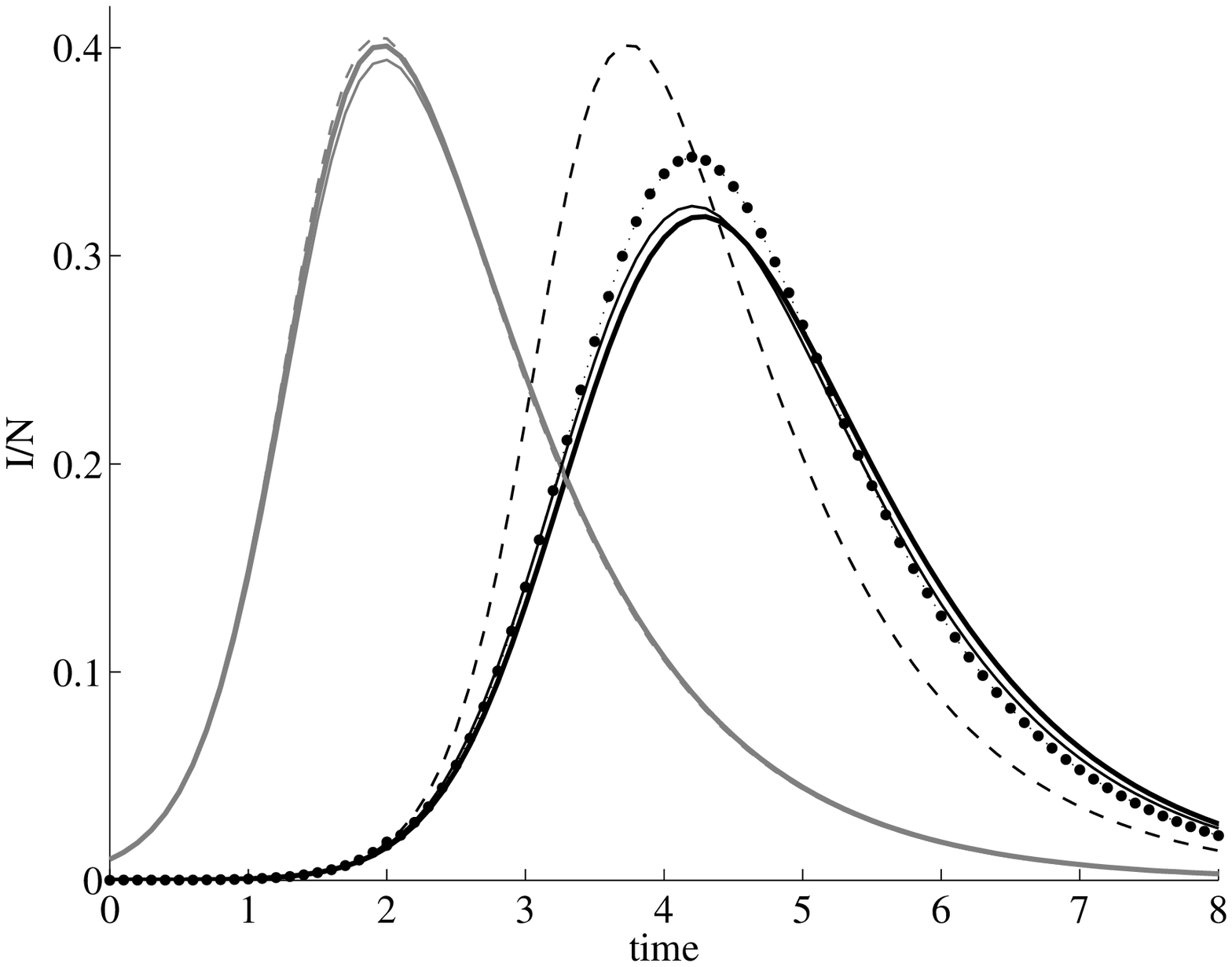} %
\includegraphics[width=60mm]{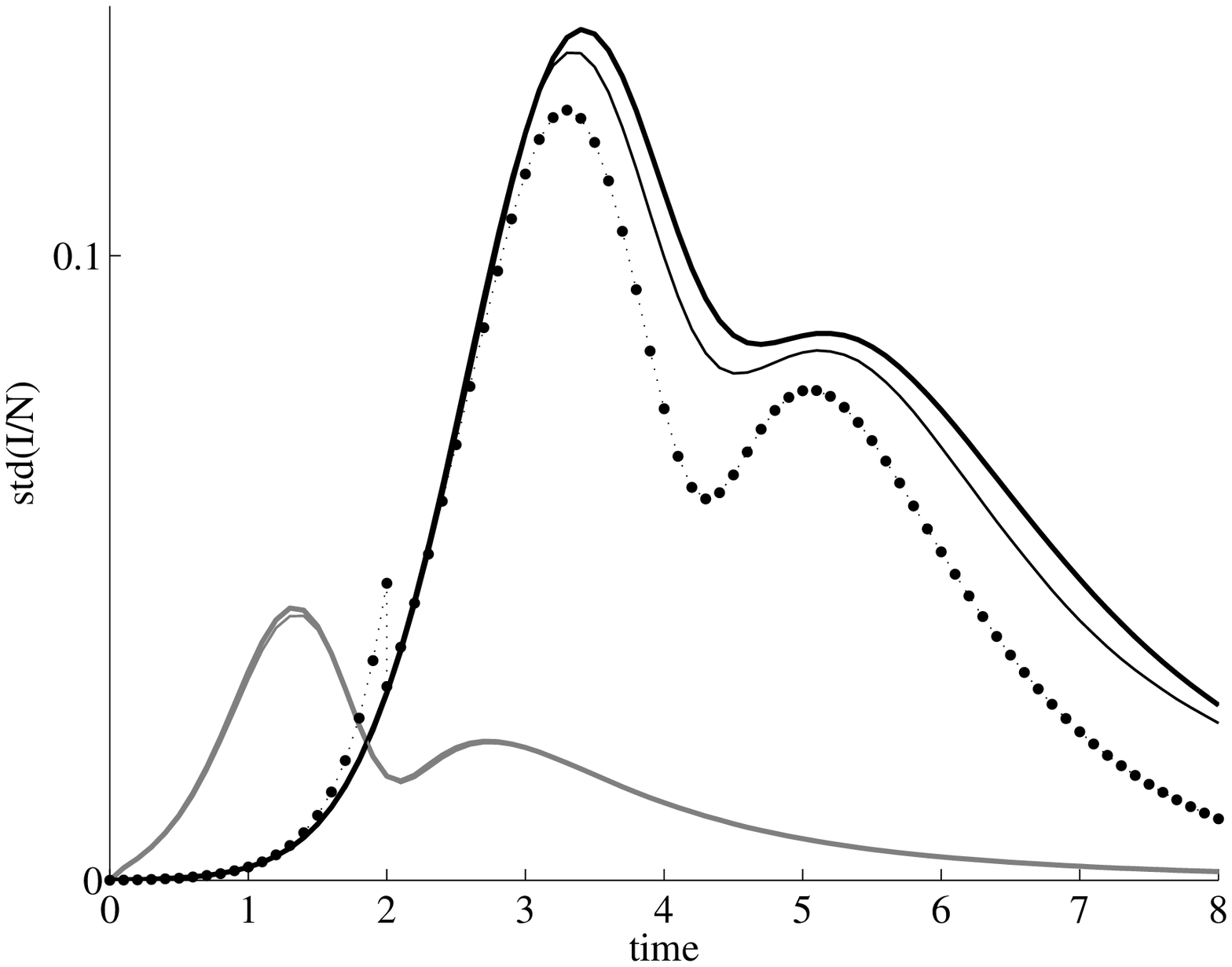}
\caption{The same as in Figure~\ref{fig:SIC10k} but for $N_1=N_2=2$k.}
\label{fig:SIC2k}
\end{figure}

Evidently, the proposed two-stage semi-randomized model gives a quite
satisfactory approximation for the first two moments of total number of
infectives in node~2 if the population is $10$k but only qualitative similarity for the smaller population.
This justifies the used simplifications for rather moderate populated sites, but for the sites with population 2k and smaller a more sophisticated models should be developed. This can be a material of the consequent works.

\section{Discussion}

The randomized network SIR model coupled by randomized migration fluxes is
described in terms of a Markov chain (MC). In the absence of infectives the
pure migration model give reasonable modelling of migration of individuals is described as a simple MC:
being disturbed the system returns to a dynamic equilibrium exponentially fast that
resembles a diffusion process in physics.

In the mean-field (hydrodynamic) limit the MC converges to a non-standard
network SIR model: the host and guest species are treated separately in the
corresponding ODEs (\ref{ODE:Sn})--(\ref{ODE:Imn}).

Note that traditionally the coupling with other nodes is described by adding
some transport terms (cf.\ \cite{DG,Wang04})
\begin{equation*}
\begin{array}{rcl}
{\frac{\mathrm{d}}{\mathrm{d}t}}S_{n} & = & -\beta _{n}S_{n}I_{n}+\chi
_{mn}^{S}S_{m} \\
{\frac{\mathrm{d}}{\mathrm{d}t}}I_{n} & = & \beta _{n}S_{n}I_{n}-\alpha
_{n}I_{n}+\chi _{mn}^{I}I_{m}%
\end{array}%
\end{equation*}
Simple analysis shows that pure migration in the such equation results in
exponentially growing solutions \cite{arXiv}. This instability is ignored as
it can be hidden in the background of the outbreak and not be observable in
certain epidemic model scenarios.

The model proposed in \cite{Wang03,Arino05}
\begin{equation*}
\begin{array}{rcl}
{\frac{\mathrm{d}}{\mathrm{d}t}}S_{n} & = & -\beta _{n}S_{n}I_{n}+\chi
_{mn}^{S}S_{m}-\chi _{nm}^{S}S_{n} \\
{\frac{\mathrm{d}}{\mathrm{d}t}}I_{n} & = & \beta _{n}S_{n}I_{n}-\alpha
_{n}I_{n}+\chi _{mn}^{I}I_{m}-\chi _{nm}^{I}I_{n}%
\end{array}%
\end{equation*}%
gives more stable pure migration but the simple analysis shows that in this
model we obtain the fully mixed population in all the nodes \cite{arXiv} ($%
\varepsilon =0.5$ in our terms). Thus the dynamics of this model seems to be
more realistic but nevertheless does not satisfy an intuitive interpretation
of the equilibrium of the migration process.

The both above models are characterized by only one parameter describing
migration of a given species. This makes impossible to tune the model to
obtain the realistic migration in the absence of the outbreak.

In our earlier work \cite{SKG11a} a migration model is introduced without
splitting species into host and guest with two migration parameters: $%
\varepsilon $ and $\tau $ describing migration process of a given species.
But model proposed in \cite{SKG11a} also does not give a completely
satisfactory solution for pure migration in the case of different migration
times: $\tau _{12}^{S}$ and $\tau _{21}^{S}$ as shown in \cite{arXiv}.

Note that for some combination of the network parameters (epidemic and
migration) effect of the more correct accounting of migration can be very
small but it can become essential when parameters of the model vary.

It is interesting to compare three different techniques for the model under
consideration: a MC describing the number of individuals from all categories
in both centers, its hydrodynamical limit in the form of a system of dynamic
equations and a simple description of contamination stage at the node 2 as
an isolated center with an inflow of infectives neglecting the backward
migration. We developed an approach when the random evolution on the
contamination stage either in its full or simplified form is coupled with
the dynamical description on the stage of saturation. This makes the problem
computationally feasible. Our intention is to apply this technique to a
network of interacting population centers in the next paper. Note that the
direct simulation of the network is far too expensive in terms of
computational time compared with the integration of systems of dynamical
equations. The computational time may be considerably reduced if the
simulation is required during a relatively short contamination periods only.

The spatial dynamics of multi-species epidemic models is widely discussed in the literature
(see \cite{Mollison91,BG1995,GKGB1998,GBK2001,Arino2007,BGM2007,CV2007,CV2008,BBCS2012} and the reference therein). For a purely deterministic model the account of different dynamics of host and guest species on the epidemic speed was studied in \cite{arXiv}. The simplifying assumptions make the analysis tractable by may not adequately reflect reality. It seems that a network of stochasticly interacting centres of the type discussed above may provide more realistic but still tractable setting.

In the next paper we intend to derive the traveling wave characteristic equation (cf. \cite{SKG08},\cite{SKG11a},\cite{arXiv}) and explore analytically an numerically the dependence of the mean epidemic speed and its standard deviation on the network parameters.


\section*{Acknowledgment}

The financial support within the framework of a subsidy granted to the National Research
University Higher School of Economics for the implementation of the Global Competitiveness is acknowledged.

\appendix

\section{Sketch of the proof to Proposition 1.}

First, the mean values of the Markov chain (MC) converges to the solution of
initial value problem (\ref{ODE:Sn})--(\ref{eq:S12:mean}) by the LLN.
Phenomenological sketch is given Section~\ref{sec:MC}, and the rigorous
proof is analogous to that presented in \cite{SW95,Manolopoulou}.

We also have to establish the convergence in probability. For definiteness
consider $I_{2}(t)$ and apply Chebyshev's inequality for any $\epsilon >0$
\begin{equation*}
\mathbb{P}\Big(\Big|\frac{I_{2}(t)}{\Lambda }-{\hat{I}}_{2}(t)\Big|>\epsilon %
\Big)\leq \frac{\mathrm{var}\left[ I_{2}(t)\right] }{\Lambda ^{2}\epsilon
^{2}}.
\end{equation*}%
Recall that $\Lambda $ is the population scaling parameter (see Section~\ref%
{sec:MC}). So, it is enough to check that
\begin{equation}
\mathrm{var}\left[ I_{2}(t)\right] =O(\Lambda ),\qquad \Lambda \rightarrow
\infty.  \label{App:cov}
\end{equation}%
It is demonstrated numerically in Section~\ref{sec:numerics} (see Figure~\ref%
{fig:stdImax-vs-N}). Actually we see that the normalized standard deviation
decays as $N^{-1/2}$, or equivalently, the non-normalized standard deviation
grows as $N^{1/2}$ (i.e., as $\Lambda^{1/2}$), that implies (\ref{App:cov}).

The rigorous argument runs as follows. Consider the processes in Table~1
which cause the change in number of infectives in node~2 and outline the
fluxes of infectives. These processes are
\begin{equation}
\begin{array}{ccc}
\hline\hline
\mbox{\#} & \mbox{Event} & \mbox{Rate}\rule[-0.5019pc]{0pc}{1.5056pc} \\
\hline\hline
\mbox{9,16} & I_{2}\rightarrow I_{2}+1 & \beta _{2}I_{2}S_{2}+\underline{%
\beta _{2}I_{12}S_{2}+\delta _{12}^{I}I_{21}}\rule[-0.5019pc]{0pc}{1.5056pc}
\\ \hline
\mbox{11,14} & I_{2}\rightarrow I_{2}-1 & \alpha _{1}I_{2}+\underline{\gamma
_{21}^{I}I_{2}}\rule[-0.5019pc]{0pc}{1.5056pc} \\ \hline\hline
\end{array}
\label{App:Process-1}
\end{equation}%
Here the flux terms are underlined, the remaining terms describe the MC
based stochastic SIR model \cite{DG,SKG11c}. So the real flux can be defined
as%
\begin{equation*}
\nu (t)=\beta _{2}I_{12}S_{2}+\delta _{12}^{I}I_{21}-\gamma _{21}^{I}I_{2}.
\end{equation*}

We construct the process $\tilde{I}_{2}$%
\begin{equation}
\begin{array}{cc}
\hline\hline
\mbox{Event} & \mbox{Rate} \rule[-0.5pc]{0pc}{1.5pc} \\ \hline\hline
\tilde{I}_{2}\rightarrow \tilde{I}_{2}+1 & \beta _{2}^{\prime }\tilde{I}_{2}+%
\tilde{\nu}\rule[-0.5pc]{0pc}{1.5pc} \\ \hline
\tilde{I}_{2}\rightarrow \tilde{I}_{2}-1 & \alpha _{1}\tilde{I}_{2}%
\rule[-0.5pc]{0pc}{1.5pc} \\ \hline\hline
\end{array}
\label{App:Process-2}
\end{equation}%
with the majorized constant flux%
\begin{equation*}
\tilde{\nu}=\beta _{2}^{\prime }N_{1}+\delta _{12}^{I}N_{2}\geq \nu (t).
\end{equation*}%
Remind that $\beta _{2}^{\prime }=\beta _{2}N_{2}$ is a constant when $%
\Lambda \rightarrow \infty $.

For this process we have a randomized SI model (considered in \cite{SKG11c})
with the constant Poisson flux $\tilde{\nu}$. This problem is solved in
Section~\ref{sec:SIC} and it is shown that its variance grows as $O(\Lambda
) $.

Next, we establish the second order stochastic domination (see \cite{MS} for
details) of process $I_{2}(t)$ by $\tilde{I}_{2}(t)$. In fact the following
inequality holds for all $x,t\geq 0$ (cf. \cite{MS})
\begin{equation*}
\int_{0}^{x}\;\mathbb{P}\big(I_{2}(t)\geq u\big)\mathrm{d}u\leq
\int_{0}^{x}\;\mathbb{P}\big({\tilde{I}}_{2}(t)\geq u\big)\mathrm{d}u.
\end{equation*}

The second order stochastic domination means that for any convex function $%
\Psi (.)$ we have the inequality for all $t\geq 0$
\begin{equation*}
\mathbb{E}[\Psi (I_{2}(t))]\leq \mathbb{E}[\Psi ({\tilde{I}}_{2}(t))],
\end{equation*}%
${\tilde{I}}_{2}(t)$ is the number of susceptible in the tractable model
described by (\ref{App:Process-2}). In our case $\Psi (X)=(X-\mathbb{E}%
X)^{2} $. This implies the inequality $\mathrm{var}\left[ I_{2}(t)\right]
\leq \mathrm{var}\left[ \tilde{I}_{2}(t)\right] $. $\Box $

\end{document}